\documentclass[sigconf]{acmart}
\usepackage{booktabs} 

\usepackage{rotating}
\usepackage{multirow}
\usepackage{graphicx}
\usepackage{caption}
\usepackage{subfig}
\usepackage[percent]{overpic}
\usepackage{rotate}
\usepackage{slashbox}
\usepackage{url}

\copyrightyear{2017}
\acmYear{2017}
\setcopyright{acmcopyright}
\acmConference{SIGIR '17}{}{August 07-11, 2017, Shinjuku, Tokyo, Japan}
\acmPrice{15.00}\acmDOI{http://dx.doi.org/10.1145/3077136.3080831}
\acmISBN{978-1-4503-5022-8/17/08}

\newcommand{\eqname}{Equation}
\newcommand{\specialcell}[2][c]{\begin{tabular}[#1]{@{}c@{}}#2\end{tabular}}

\newcommand\topspace{\rule{0pt}{2.6ex}}       

\newcommand{\firstmodel}{RLM~}
\newcommand{\secondmodel}{RPE~}

\newcommand{\firstmodelc}{Relevance Likelihood Maximization Model}
\newcommand{\secondmodelc}{Relevance Posterior Estimation Model}

\begin{document}

\clubpenalty=10000 
\widowpenalty = 10000


\title{Relevance-based Word Embedding}

\author{Hamed Zamani}
\affiliation{%
  \institution{Center for Intelligent Information Retrieval}
  \institution{College of Information and Computer Sciences}
  \institution{University of Massachusetts Amherst}
  \city{Amherst} 
  \state{MA} 
  \postcode{01003}
}
\email{zamani@cs.umass.edu}

\author{W. Bruce Croft}
\affiliation{%
  \institution{Center for Intelligent Information Retrieval}
  \institution{College of Information and Computer Sciences}
  \institution{University of Massachusetts Amherst}
  \city{Amherst} 
  \state{MA} 
  \postcode{01003}
}
\email{croft@cs.umass.edu}



\begin{abstract}
Learning a high-dimensional dense representation for vocabulary terms, also known as a word embedding, has recently attracted much attention in natural language processing and information retrieval tasks. The embedding vectors are typically learned based on term proximity in a large corpus. This means that the objective in well-known word embedding algorithms, e.g., word2vec, is to accurately predict adjacent word(s) for a given word or context. However, this objective is not necessarily equivalent to the goal of many information retrieval (IR) tasks. The primary objective in various IR tasks is to capture \textit{relevance} instead of term proximity, syntactic, or even semantic similarity. This is the motivation for developing unsupervised relevance-based word embedding models that learn word representations based on query-document relevance information. In this paper, we propose two learning models with different objective functions; one learns a relevance distribution over the vocabulary set for each query, and the other classifies each term as belonging to the relevant or non-relevant class for each query. To train our models, we used over six million unique queries and the top ranked documents retrieved in response to each query, which are assumed to be relevant to the query. We extrinsically evaluate our learned word representation models using two IR tasks: query expansion and query classification. Both query expansion experiments on four TREC collections and query classification experiments on the KDD Cup 2005 dataset suggest that the relevance-based word embedding models significantly outperform state-of-the-art proximity-based embedding models, such as word2vec and GloVe.

\end{abstract}





\maketitle


\section{Introduction}
\label{sec:intro}
Representation learning is a long-standing problem in natural language processing (NLP) and information retrieval (IR). The main motivation is to abstract away from the surface forms of a piece of text, e.g., words, sentences, and documents, in order to alleviate sparsity and learn meaningful similarities, e.g., semantic or syntactic similarities, between two different pieces of text. Learning representations for words as the atomic components of a language, also known as word embedding, has recently attracted much attention in the NLP and IR communities.

A popular model for learning word representation is neural network-based language models. For instance, the word2vec model proposed by Mikolov et al.~\cite{Mikolov:2013} is an embedding model that learns word vectors via a neural network with a single hidden layer. Continuous bag of words (CBOW) and skip-gram are two implementations of the word2vec model. Another successful trend in learning semantic word representations is employing global matrix factorization over word-word matrices. GloVe~\cite{Pennington:2014} is an example of such methods. A theoretical relation has been discovered between embedding models based on neural network and matrix factorization in \cite{Levy:2014}. These models have been demonstrated to be effective in a number of IR tasks, including query expansion \cite{Diaz:2016,Kuzi:2016,Zamani:2016:ICTIR:emb}, query classification \cite{Liu:2015,Zamani:2016:ICTIR:pqv}, short text similarity \cite{Kenter:2015}, and document model estimation \cite{Ai:2016,Rekabsaz:2016}.

The aforementioned embedding models are typically trained based on term proximity in a large corpus. For instance, the word2vec model's objective is to predict adjacent word(s) given a word or context, i.e., a context window around the target word. This idea aims to capture semantic and syntactic similarities between terms, since semantically/syntactically similar words often share similar contexts. However, this objective is not necessarily equivalent to the main objective of many IR tasks. The primary objective in many IR methods is to model the notion of \textit{relevance}~\cite{Lavrenko:2001,Saracevic:2016,Zhai:2003}. In this paper, we revisit the underlying assumption of typical word embedding methods, as follows:

\smallskip
\textit{The objective is to predict the words observed in the documents relevant to a particular information need.}
\smallskip

This objective has been previously considered for developing relevance models \cite{Lavrenko:2001}, a state-of-the-art (pseudo-) relevance feedback approach. Relevance models try to optimize this objective given a set of relevant documents for a given query as the indicator of user's information need. In the absence of relevance information, the top ranked documents retrieved in response to the query are assumed to be relevant. Therefore, relevance models, and in general all pseudo-relevance feedback models, use an online setting to obtain training data: retrieving documents for the query and then using the top retrieved documents in order to estimate the relevance distribution. Although relevance models have been proved to be effective in many IR tasks \cite{Lavrenko:2001,Lavrenko:2002}, having a retrieval run for each query to obtain the training data for estimating the relevance distribution is not always practical in real-world search engines. We, in this paper, optimize a similar objective in an offline setting, which enables us to predict the relevance distribution without any retrieval runs during the test time. To do so, we consider the top retrieved documents for millions of training queries as a training set and learn embedding vectors for each term in order to predict the words observed in the top retrieved documents for each query. We develop two relevance-based word embedding models. The first one, the relevance likelihood maximization model (RLM), aims to model the relevance distribution over the vocabulary terms for each query, while the second one, the relevance posterior estimation model (RPE), classifies each term as  relevant or non-relevant to each query. We provide efficient learning algorithms to train these models on large amounts of training data. Note that our models are unsupervised and the training data is generated automatically.

To evaluate our models, we performed two sets of extrinsic evaluations. In the first set, we focus on the query expansion task for ad-hoc retrieval. In this set of experiments, we consider four TREC collections, including two newswire collections (AP and Robust) and two large-scale web collections (GOV2 and ClueWeb09 - Cat. B). Our results suggest that the relevance-based embedding models outperform state-of-the-art word embedding algorithms. The \firstmodel model shows better performance compared to \secondmodel in the context of query expansion, since the goal is to estimate the probability of each term given a query and this distribution is not directly learned by the \secondmodel model. In the second set of experiments, we focus on the query classification task using the KDD Cup 2005 \cite{Li:2005} dataset. In this extrinsic evaluation, the relevance-based embedding models again perform better than the baselines. Interestingly, the query classification results demonstrate that the \secondmodel model outperforms the \firstmodel model, for the reason that in this task, unlike the query expansion task, the goal is to compute the similarity between two query vectors, and \secondmodel can learn more accurate embedding vectors with less training data.

\vspace{-0.4cm}
\section{Related Work}
\label{sec:rel}
Learning a semantic representation for text has been studied for many years. Latent semantic indexing (LSI) \cite{Deerwester:1988} can be considered as early work in this area that tries to map each text to a semantic space using singular value decomposition (SVD), a well-known matrix factorization algorithm. Subsequently, \citet{Clinchant:2013} proposed Fisher Vector (FV), a document representation framework based on continuous word embeddings, which aggregates a non-linear mapping of word vectors into a document-level representation. However, a number of popular IR models, such as BM25 and language models, often significantly outperform the models that are based on semantic similarities. Recently, extremely efficient word embedding algorithms have been proposed to model semantic similarly between words.

Word embedding, also known as distributed representation of words, refers to a set of machine learning algorithms that learn high-dimensional real-valued dense vector representation $\vec{w} \in \mathbb{R}^d$ for each vocabulary term $w$, where $d$ denotes the embedding dimensionality. GloVe \cite{Pennington:2014} and word2vec \cite{Mikolov:2013} are two well-known word embedding algorithms that learn embedding vectors based on the same idea, but using different machine learning techniques. The idea is that the words that often appear in similar contexts are similar to each other. To do so, these algorithms try to accurately predict the adjacent word(s) given a word or a context (i.e., a few words appeared in the same context window). Recently, Rekabsaz et al. \cite{Rekabsaz:2017} proposed to exploit global context in word embeddings in order to avoid topic shifting.

Word embedding representations can be also learned as a set of parameters in an end-to-end neural network model. For instance, Zamani et al. \cite{Zamani:2017} trained a context-aware ranking model in which the embedding vectors of frequent n-grams are learned using click data. More recently, Dehghani et al. \cite{Dehghani:2017} trained neural ranking models with weak supervision data (i.e., a set of noisy training data automatically generated by an existing unsupervised model) that learn word representations in an end-to-end ranking scenario.

Word embedding vectors have been successfully employed in several NLP and IR tasks. Kusner et al.~\cite{Kusner:2015} proposed word mover's distance (WMD), a function for calculating semantic distance between two documents, which measures the minimum traveling distance from the embedded vectors of individual words in one document to the other one. 
Zhou et al.~\cite{Zhou:2015} introduced an embedding-based method for question retrieval in the context of community question answering. \citet{Vulic:2015} proposed a model to learn bilingual word embedding vectors from document-aligned comparable corpora. Zheng and Callan~\cite{Zheng:2015} presented a supervised embedding-based technique to re-weight terms in the existing IR models, e.g., BM25. Based on the well-defined structure of language modeling framework in information retrieval, a number of methods have been introduced to employ word embedding vectors within this framework in order to improve the performance in IR tasks. For instance, 
Zamani and Croft \cite{Zamani:2016:ICTIR:emb} presented a set of embedding-based query language models using the query expansion and pseudo-relevance feedback techniques that benefit from the word embedding vectors. Query expansion using word embedding has been also studied in \cite{Diaz:2016,Kuzi:2016,Sordoni:2014}. All of these approaches are based on word embeddings learned based on term proximity information. PhraseFinder \cite{Jing:1994} is an early work using term proximity information for query expansion. Mapping vocabulary terms to HAL space, a low-dimensional space compared to vocabulary size, has been used in \cite{Bruza:2002} for query modeling.


As is widely known in the information retrieval literature \cite{Diaz:2016,Xu:1996}, there is a big difference between the unigram distribution of words on sub-topics of a collection and the unigram distribution estimated from the whole collection. Given this phenomenon, Diaz et al. \cite{Diaz:2016} recently proposed to train word embedding vectors on the top retrieved documents for each query. However, this model, called local embedding, is not always practical in real-word applications, since the embedding vectors need to be trained during the query time. Furthermore, the objective function in local embedding is based on term proximity in pseudo-relevant documents. 

In this paper, we propose two models for learning word embedding vectors, that are specifically designed for information retrieval needs. All the aforementioned tasks in this section can potentially benefit from the vectors learned by the proposed models.

\vspace{-0.3cm}
\section{Relevance-based Embedding}
\label{sec:method}

Typical word embedding algorithms, such as word2vec \cite{Mikolov:2013} and GloVe \cite{Pennington:2014}, learn high-dimensional real-valued embedding vectors based on the proximity of terms in a training corpus, i.e., co-occurrence of terms in the same context window. Although these approaches could be useful for learning the embedding vectors that can capture semantic and syntactic similarities between vocabulary terms and have shown to be useful in many NLP and IR tasks, there is a large gap between their learning objective (i.e., term proximity) and what is needed in many information retrieval tasks. For example, consider the query expansion task and assume that a user submitted the query ``dangerous vehicles''. One of the most similar terms to this query based on the typical word embedding algorithms (e.g., word2vec and GloVe) is ``safe'', and thus it would get a high weight in the expanded query model. The reason is that the words ``dangerous'' and ``safe'' often share similar contexts. However, expanding the query with the word ``safe'' could lead to poor retrieval performance, since it changes the meaning and the intent of the query.

This example together with many others have motivated us to revisit the objective used in the learning process of word embedding algorithms in order to obtain the word vectors that better match with the needs in IR tasks. The primary objective in many IR tasks is to model the notion of \textit{relevance}. Several approaches, such as the relevance models proposed by Lavrenko and Croft \cite{Lavrenko:2001}, have been proposed to model relevance. Given the successes achieved by these models, we propose to learn word embedding vectors based on an objective that matters in information retrieval. The objective is to accurately predict the terms that are observed in a set of relevant documents to a particular information need. 

In the following subsections, we first describe our neural network architecture, and then explain how to build a training set for learning relevance-based word embeddings. We further introduce two models, relevance likelihood maximization (RLM) and relevance posterior estimation (RPE), with different objectives using the described neural network.


\subsection{Neural Network Architecture}
\label{sec:nn}
\begin{figure}
    \centering
    \vspace{-0.2cm}
    \includegraphics[clip, trim=0cm 3cm 4cm 1cm, width=0.8\textwidth]{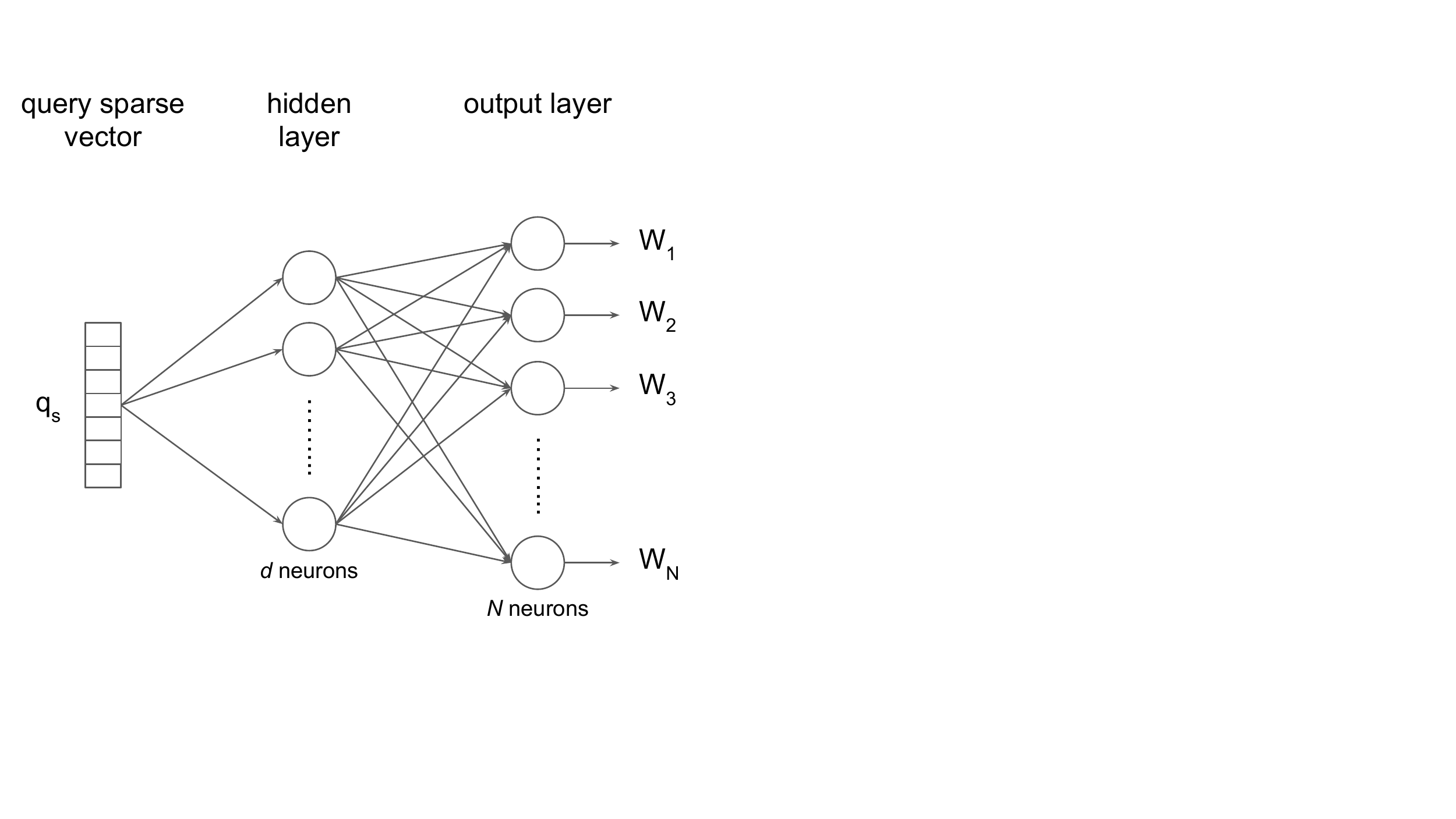}
    \caption{The relevance-based word embedding architecture. The objective is to learn $d$-dimensional distributed representation for words based on the notion of relevance, instead of term proximity. $N$ denotes the total number of vocabulary terms.}
    \vspace{-0.3cm}
    \label{fig:arch}
\end{figure}

We use a simple yet effective feed-forward neural network with a single linear hidden layer. The architecture of our neural network is shown in \figurename~\ref{fig:arch}. The input of the model is a sparse query vector $\vec{q}_s$ with the length of $N$, where $N$ denotes the total number of vocabulary terms. This vector can be obtained by a projection function given the vectors corresponding to individual query terms. In this paper, we simply consider average as the projection function. Hence, $\vec{q}_s = \frac{1}{|q|} \sum_{w \in q}{\vec{e}_w}$, where $\vec{e}_w$ and $|q|$ denote the one-hot vector representation of term $w$ and the query length, respectively. The hidden layer in this network maps the given query sparse vector to a query embedding vector $\vec{q}$, as follows:
\begin{equation}
    \vec{q} = \vec{q}_s \times \mathcal{W}_Q
\end{equation}
where $\mathcal{W}_Q \in \mathbb{R}^{N \times d}$ is a weight matrix for estimating query embedding vectors and $d$ denotes the embedding dimensionality. The output layer of the network is a fully-connected layer given by:
\begin{equation}
    \sigma(\vec{q} \times \mathcal{W}_w + b_w)
    \label{eq:hidden_layer}
\end{equation}
where $\mathcal{W}_w \in \mathbb{R}^{d \times N}$ and $b_w \in \mathbb{R}^{1 \times N}$ are the weight and the bias matrices for estimating the probability of each term. $\sigma$ is the activation function which is discussed in Sections~\ref{sec:model1} and \ref{sec:model2}.

To summarize, our network contains two sets of embedding parameters, $\mathcal{W}_Q$ and $\mathcal{W}_w$. The former aims to map the query into the ``query embedding space'', while the latter is used to estimate the weights of individual terms. 

\subsection{Modeling Relevance for Training}
\label{sec:prf}
Relevance feedback has been shown to be highly effective in improving retrieval performance \cite{Rocchio:1971,Croft:2009}. In relevance feedback, a set of relevant documents to a given query is considered for estimating accurate query models. Since explicit relevance signals for a given query are not always available, pseudo-relevance feedback (PRF) assumes that the top retrieved documents in response to the given query are relevant to the query and uses these documents in order to estimate better query models. The effectiveness of PRF in various retrieval scenarios indicates that useful information can be captured from the top retrieved documents \cite{Lavrenko:2001,Lavrenko:2002,Zhai:2001}. In this paper, we make use of this well-known assumption to train our model. It should be noted that there is a significant difference between PRF and the proposed models: In PRF, the feedback model is estimated from the top retrieved documents of the given query in an online setting. In other words, PRF retrieves the documents for the initial query and then estimates the feedback model using the top retrieved documents. In this paper, we propose to train the model in an offline setting. Moving from the online to the offline setting would lead to substantial improvements in efficiency, because an extra retrieval run is not needed in the offline setting. To learn a model in an offline setting, we consider a fixed-length dense vector for each vocabulary term and estimate these vectors based on the information extracted from the top retrieved documents for large numbers of training queries. Note that our models are unsupervised. However, if explicit relevance data is available, such as click data, without loss of generality, both the explicit or implicit relevant documents can be considered for training our models. We leave studying the vectors learned based on supervised signals for future work.


To formally describe our training data, let $T=\{(q_1, \mathcal{R}_1), (q_2, \mathcal{R}_2),\\ \cdots, (q_m, \mathcal{R}_m)\}$ be a training set with $m$ training queries. The $i^{th}$ element of this set is a pair of query $q_i$ and the corresponding pseudo-relevance feedback distribution. These distributions are estimated based on the top $k$ retrieved documents (in our experiments, we set $k$ to $10$) for each query. The distributions can be estimated using any PRF model, such as those proposed in \cite{Lavrenko:2001,Tao:2006b,Zamani:2016:CIKM,Zhai:2001}. In this paper, we only focus on the relevance model \cite{Lavrenko:2001}, a state-of-the-art PRF model, that estimates the relevance distribution as:
\begin{equation}
    p(w|\mathcal{R}_i) \propto \sum_{d \in F_i}{p(w|d) \prod_{w' \in q_i}{p(w' | d)}}
\end{equation}
where $F_i$ denotes a set of top retrieved documents for query $q_i$. Note that the probability of terms that do not appear in the top retrieved documents is equal to zero.

\subsection{\firstmodelc}
\label{sec:model1}
In this model, the goal is to learn the relevance distribution $\mathcal{R}$. Given a set of training data, we aim to find a set of parameters $\theta_\mathcal{R}$ in order to maximize the likelihood of generating relevance model probabilities for the whole training set. The likelihood function is defined as follows:
\begin{equation}
    \prod_{i=1}^{m}{\prod_{w \in V_i}{\widehat{p}(w | q_i ; \theta_\mathcal{R})^{p(w | \mathcal{R}_i)}}}
\end{equation}
where $\widehat{p}$ is the relevance distribution that can be obtained given the learning parameters $\theta_\mathcal{R}$ and $p(w | \mathcal{R}_i)$ denotes the relevance model distribution estimated for the $i^{th}$ query in the training set (see Section~\ref{sec:prf} for more detail). $V_i$ denotes a subset of vocabulary terms that appeared in the top ranked documents retrieved for the query $q_i$. The reason for iterating over the terms that appeared in this set instead of the whole vocabulary set $V$ is that the probability $p(w | \mathcal{R}_i)$  is equal to zero for all terms $w \in V-V_i$.

In this method, we model the probability distribution $\widehat{p}$ using the softmax function (i.e., the function $\sigma$ in \eqname~\eqref{eq:hidden_layer}) as follows:\footnote{For simplicity, we drop the bias term in these equations.}
\begin{equation}
    \widehat{p}(w | q ; \theta_\mathcal{R}) = \frac{\exp{(\vec{w}^T \vec{q})}}{\sum_{w' \in V}{\exp{(\vec{w'}^T \vec{q})}}}
    \label{eq:model1:prob}
\end{equation}
where $\vec{w}$ denotes the learned embedding vector for term $w$ and $\vec{q}$ is the query vector came from the output of the hidden layer in our network (see Section~\ref{sec:nn}). According to the softmax modeling and the log-likelihood function, we have the following objective:
\begin{equation}
    \arg\max_{\theta_\mathcal{R}} \sum_{i=1}^{m}{\sum_{w \in V_i}{p(w | \mathcal{R}_i) \left(\log \exp{(\vec{w}^T \vec{q_i})} - \log \sum_{w' \in V}{\exp{(\vec{w'}^T \vec{q_i})}} \right)}}
\end{equation}

Computing this objective function and its derivatives would be computationally expensive (due to the presence of the normalization factor $\sum_{w' \in V}{\exp{(\vec{w'}^T \vec{q})}}$ in the objective function). Since all the word embedding vectors as well as the query vector are changed during the optimization process, we cannot simply omit the normalization term as is done in \cite{Zamani:2016:ICTIR:pqv} for estimating query embedding vectors based on pre-trained word embedding vectors. To make the computations more tractable, we consider a hierarchical approximation of the softmax function, which was introduced by Morin and Bengio \cite{Morin:2005} in the context of neural network language models and then successfully employed by Mikolov et al. \cite{Mikolov:2013} in the word2vec model.

The hierarchical softmax approximation uses a binary tree structure to represent the vocabulary terms, where each leaf corresponds to a unique word. There exists a unique path from the root to each leaf, and this path is used for estimating the probability of the word representing by the leaf. Therefore, the complexity of calculating softmax probabilities goes down from $O(|V|)$ to $O(\log(|V|))$ which is the height of the tree. This leads to a huge improvement in computational complexity. We refer the reader to \cite{Morin:2005,Minh:2009} for the details of calculating the hierarchical softmax approximation.

\subsection{\secondmodelc}
\label{sec:model2}
As an alternative to maximum likelihood estimation, we can estimate the relevance posterior probability. In the context of pseudo-relevance feedback, Zhai and Laffery \cite{Zhai:2001} assumed that the language model of the top retrieved documents is estimated based on a mixture model. In other words, it is assumed that there are two language models for the feedback set: the relevance language model\footnote{The phrase ``topical language model'' was used in the original work \cite{Zhai:2001}. We call it ``relevance language model'' to have consistent definitions in our both models.} and a background noisy language model. They used an expectation-maximization algorithm to estimate the relevance language model. In this model, we make use of this assumption in order to cast the problem of estimating the relevance distribution $\mathcal{R}$ as a classification task: \textit{Given a pair of word $w$ and query $q$, does $w$ come from the relevance distribution of the query $q$?} Instead of $p(w|\mathcal{R})$, this model estimates $p(R=1 | w, q; \theta_\mathcal{R})$ where $R$ is a Boolean variable and $R=1$ means that the given term-query pair $(w, q)$ comes from the relevance distribution $\mathcal{R}$. $\theta_\mathcal{R}$ is a set of parameters that is going to be learned during the training phase.

Therefore, the problem is cast as a binary classification task that can be modeled by logistic regression (which means the function $\sigma$ in \eqname~\eqref{eq:hidden_layer} is the sigmoid function):
\begin{equation}
    \widehat{p}(R=1 | \vec{w}, \vec{q}; \theta_\mathcal{R}) = \frac{1}{1 + e^{(-\vec{w}^T \vec{q})}}
    \label{eq:model2:prob}
\end{equation}
where $\vec{w}$ is the relevance-based word embedding vector for term $w$. Similar to the previous model, $\vec{q}$ is the output of the hidden layer of the network, representing the query embedding vector.

In order to address this binary classification problem, we consider a cross-entropy loss function. In theory, for each training query, our model should learn to model relevance for the terms appearing in the corresponding pseudo-relevant set and non-relevance for all the other vocabulary terms, which could be impractical, due to the large number of vocabulary terms. Similar to \cite{Mikolov:2013}, we propose to use the noise contrastive estimation (NCE) \cite{Gutmann:2012} which hypothesizes that we can achieve a good model by only differentiating the data from noise via a logistic regression model. The main concept in NCE is similar to those proposed in the divergence from randomness model \cite{Amati:2002} and the divergence minimization feedback model \cite{Zhai:2001}. Based on the NCE hypothesis, we define the following negative cross-entropy objective function for training our model:
\begin{align}
    \arg\max_{\theta_\mathcal{R}} \sum_{i=1}^{m}&\left[{\sum_{j=1}^{\eta^+}{\mathbb{E}_{w_j \sim p(w | \mathcal{R}_i)} \left[\log{\widehat{p}(R=1 | \vec{w_j}, \vec{q_i}; \theta_\mathcal{R})}\right]}}\right. \nonumber\\
    &+ \left. \sum_{j=1}^{\eta^-}{\mathbb{E}_{w_j \sim p_n(w)} \left[\log{\widehat{p}(R=0 | \vec{w_j}, \vec{q_i}; \theta_\mathcal{R})}\right]}\right]
\end{align}
where $p_n(w)$ denotes a noise distribution and $\eta = (\eta^+, \eta^-)$ is a pair of hyper-parameters to control the number of positive and negative instances per query, respectively. We can easily calculate $\widehat{p}(R=0 | \vec{w_j}, \vec{q_i}) = 1 - \widehat{p}(R=1 | \vec{w_j}, \vec{q_i})$. The noise distribution $p_n(w)$ can be estimated using a function of unigram distribution $U(w)$ in the whole training set. Similar to \cite{Mikolov:2013}, we use $p_n(w) \propto U(w)^{3/4}$ which has been empirically shown to work effectively for negative sampling.

It is notable that although this model learns embedding vectors for both queries and words, it is not obvious how to calculate the probability of each term given a query; because \eqname~\ref{eq:model2:prob} only gives us a classification probability and we cannot simply use the Bayes rule here (since, not all probability components are known). This model can perform well when computing the similarity between two terms or two queries, but not a query and a term. However, we can use the model presented in \cite{Zamani:2016:ICTIR:pqv} to estimate the query model using the word embedding vectors (not the ones learned for query vectors) and then calculate the similarity between a query and a term.

\section{Experiments}
\label{sec:exp}
In this section, we first describe how we train the relevance-based word embedding models. We further extrinsically evaluate the learned embeddings using two IR tasks: query expansion and query classification. Note that the main aim here is to compare the proposed models with the existing word embedding algorithms, not with the state-of-the-art query expansion and query classification models.

\subsection{Training}
In order to train relevance-based word embeddings, we obtained millions of unique queries from the publicly available AOL query logs \cite{Pass:2006}. This dataset contains a sample of web search queries from real users submitted to the AOL search engine within a three-month period from March 1, 2006 to May 31, 2006. We only used query strings and no session and click information was obtained from this dataset. We filtered out the navigational queries containing URL substrings, i.e., ``http'', ``www.'', ``.com'', ``.net'', ``.org'', ``.edu''. All non-alphanumeric characters were removed from all queries. 
Applying all these constraints leads to over $6$ millions unique queries as our training query set. To estimate the relevance model distributions in the training set, we considered top $10$ retrieved documents in a target collection in response to each query using the Galago\footnote{\url{http://www.lemurproject.org/galago.php}} implementation of the query likelihood retrieval model \cite{Ponte:1998} with Dirichlet prior smoothing ($\mu=1500$) \cite{Zhai:2004}.

We implemented and trained our models using TensorFlow\footnote{\url{http://tensorflow.org/}}. The networks are trained based on the stochastic gradient descent optimizer using the back-propagation algorithm \cite{Rumelhart:1986} to compute the gradients. All model hyper-parameters were tuned on the training set (the hyper-parameters with the smallest training loss value were selected). For each model, the learning rate and the batch size were selected from $[0.001, 0.01, 0.1, 1]$ and $[64, 128, 256]$, respectively. For \secondmodel, we also tuned the number of positive and negative instances (i.e., $\eta^+$ and $\eta^-$). The value of $\eta^+$ was swept between $[20, 50, 100, 200]$ and the parameter $\eta^-$ was selected from $[5\eta^+, 10\eta^+, 20\eta^+]$. As suggested in \cite{Zamani:2016:ICTIR:emb}, in all the experiments (unless otherwise stated) the embedding dimensionality was set to $300$, for all models including the baselines.

\subsection{Evaluation via Query Expansion}
In this subsection, we evaluate the embedding models in the context of query expansion for the ad-hoc retrieval task. In the following, we first describe the retrieval collections used in our experiments. We further explain our experimental setup as well as the evaluation metrics. We finally report and discuss the query expansion results.

\begin{table*}
\centering
\caption{Collections statistics.}\vspace{-0.25cm}
\begin{tabular}{|c|c|c|c|c|c|} \hline
ID & collection & queries (title only) & \#docs & avg doc length & \#qrels \\ \hline
AP & Associated Press 88-89 & TREC 1-3 Ad-Hoc Track, topics 51-200 & 165k & 287 & 15,838 \\ \hline
Robust & \specialcell{TREC Disks 4 \& 5 minus \\ Congressional Record} & \specialcell{TREC 2004 Robust Track, \\ topics 301-450 \& 601-700} & 528k & 254  & 17,412 \\ \hline
GOV2 & \specialcell{2004 crawl of .gov domains} & \specialcell{TREC 2004-2006 Terabyte Track,\\topics 701-850} & 25m  & 648 & 26,917  \\ \hline
ClueWeb & \specialcell{ClueWeb 09 - Category B} & \specialcell{TREC 2009-2012 Web Track\\topics 1-200} & 50m  & 1506 & 18,771  \\ \hline
\end{tabular}
\label{tab:dataset}
\end{table*}

\begin{table*}[t]
\centering
\caption{Evaluating relevance-based word embeddings in the context of query expansion. The superscripts 0/1/2/3/4 denote that the MAP improvements over MLE/word2vec-external/word2vec-target/GloVe-external/GloVe-target are statistically significant. The highest value in each row is marked in bold.}
\vspace{-0.2cm}
\begin{tabular}{|c|l|c|c|c|c|c||l|l|} \hline
\multirow{2}{*}{Collection} & \multicolumn{1}{c|}{\multirow{2}{*}{Metric}} & \multirow{2}{*}{MLE} & \multicolumn{2}{c|}{word2vec} & \multicolumn{2}{c||}{GloVe} & \multicolumn{2}{c|}{Rel.-based Embedding} \topspace 
\\ \cline{4-9}
 &  &  & external & target & external & target & \multicolumn{1}{c|}{\firstmodel} & \multicolumn{1}{c|}{\secondmodel} \topspace 
\\ \hline

\multirow{3}{*}{AP}
& MAP     & 0.2197 & 0.2399 & 0.2420 & 0.2319 & 0.2389 & \textbf{0.2580}\textsuperscript{01234} & 0.2543\textsuperscript{01234} \topspace\\
& P@20    & 0.3503 & 0.3688 & 0.3738 & 0.3581 & 0.3631 & \textbf{0.3886}\textsuperscript{01234} & 0.3812\textsuperscript{034} \\
& NDCG@20 & 0.3924 & 0.4030 & 0.4181 & 0.4025 & 0.4098 & \textbf{0.4242}\textsuperscript{01234} & 0.4226\textsuperscript{01234} \\
\hline

\multirow{3}{*}{Robust}
& MAP     & 0.2149 & 0.2218 & 0.2215 & 0.2209 & 0.2172 & \textbf{0.2450}\textsuperscript{01234} & 0.2372\textsuperscript{01234} \topspace\\
& P@20    & 0.3319 & 0.3357 & 0.3337 & 0.3345 & 0.3281 & \textbf{0.3476}\textsuperscript{01234} & 0.3409\textsuperscript{024} \\
& NDCG@20 & 0.3863 & 0.3918 & 0.3881 & 0.3918 & 0.3844 & \textbf{0.3982}\textsuperscript{01234} & 0.3955\textsuperscript{0} \\
\hline



\multirow{3}{*}{GOV2}
& MAP     & 0.2702 & 0.2740 & 0.2723 & 0.2718 & 0.2709 & \textbf{0.2867}\textsuperscript{01234} & 0.2855\textsuperscript{01234} \topspace\\
& P@20    & 0.5132 & 0.5257 & 0.5172 & 0.5186 & 0.5128 & \textbf{0.5367}\textsuperscript{01234} & 0.5358\textsuperscript{01234} \\
& NDCG@20 & 0.4482 & 0.4571 & 0.4509 & 0.4539 & 0.4485 & \textbf{0.4576}\textsuperscript{0234} & 0.4557\textsuperscript{024} \\
\hline

\multirow{3}{*}{ClueWeb}
& MAP     & 0.1028 & 0.1033 & 0.1033 & 0.1029 & 0.1026 & \textbf{0.1066}\textsuperscript{01234} & 0.1031 \topspace\\
& P@20    & 0.3025 & 0.3040 & 0.3053 & 0.3033 & 0.3048 & \textbf{0.3073} & 0.3030 \\
& NDCG@20 & 0.2237 & 0.2235 & 0.2252 & 0.2244 & 0.2244 & \textbf{0.2273}\textsuperscript{01} & 0.2241 \\
\hline

\end{tabular}
\label{tab:main}
\vspace{-0.2cm}
\end{table*}

\subsubsection{Data}
We use four standard test collections in our experiments. The first two collections (AP and Robust) consist of thousands of news articles and are considered as homogeneous collections. AP and Robust were previously used in TREC 1-3 Ad-Hoc Track and TREC 2004 Robust Track, respectively. The second two collections (GOV2 and ClueWeb) are large-scale web collections containing heterogeneous documents. GOV2 consists of the ``.gov'' domain web pages, crawled in 2004. ClueWeb (i.e., ClueWeb09-Category B) is a common web crawl collection that only contains English web pages. GOV2 and ClueWeb were previously used in TREC 2004-2006 Terabyte Track and TREC 2009-2012 Web Track, respectively. The statistics of these collections as well as the corresponding TREC topics are reported in \tablename~\ref{tab:dataset}. We only used the title of topics as queries.

\subsubsection{Experimental Setup}
We cleaned the ClueWeb collection by filtering out the spam documents. The spam filtering phase was done using the Waterloo spam scorer\footnote{\url{http://plg.uwaterloo.ca/~gvcormac/clueweb09spam/}} \cite{Cormack:2011} with the threshold of $60\%$. Stopwords were removed from all collections using the standard INQUERY stopword list and no stemming were performed.

For the purpose of query expansion, we consider the language modeling framework \cite{Ponte:1998} and estimate a query language model based on a given set of word embedding vectors. The expanded query language model $p(w | \theta^*_q)$ is estimated as:
\begin{equation}
    p(w | \theta^*_q) = \alpha p_{ML}(w | q) + (1-\alpha) p(\vec{w} | \vec{q})
    \label{eq:expansion}
\end{equation}
where $p_{ML}(w | q)$ denotes maximum likelihood estimation of the original query and $\alpha$ is a free hyper-parameter that controls the weight of original query model in the expanded model. The probability $p(\vec{w} | \vec{q})$ is calculated based on the trained word embedding vectors. In our first model, this probability can be estimated using \eqname~\eqref{eq:model1:prob}; while in the second model, we should simply use the Bayes rule given \eqname~\eqref{eq:model2:prob} to estimate this probability. However, since we do not have any information about the probability of each term given a query, we use the uniform distribution. For other word embedding models (i.e., word2vec and GloVe), we use the standard method described in \cite{Diaz:2016}. For all the models, we ignore the terms whose embedding vectors are not available.

We retrieve the documents for the expanded query language model using the KL-divergence formula \cite{Lafferty:2001} with Dirichlet prior smoothing ($\mu=1500$) \cite{Zhai:2004}. All the retrieval experiments were carried out using the Galago toolkit \cite{Croft:2009}.

In all the experiments, the parameters $\alpha$ (the linear interpolation coefficient) and $m$ (the number of expansion terms) were set using 2-fold cross-validation over the queries in each collection. We selected the parameter $\alpha$ from $\{0.1, \dots, 0.9\}$ and the parameter $m$ from $\{10, 20, ..., 100\}$.

\subsubsection{Evaluation Metrics}
To evaluate the effectiveness of query expansion models, we report three standard evaluation metrics: mean average precision (MAP) of the top ranked $1000$ documents, precision of the top $20$ retrieved documents (P@20), and normalized discounted cumulative gain \cite{Jarvelin:2002} calculated for the top $20$ retrieved documents (nDCG@20). Statistically significant differences of MAP, P@20, and nDCG@20 values based on the two-tailed paired t-test are computed at a $95\%$ confidence level (i.e., $p\_value<0.05$).

\begin{table*}[t]
\centering
\caption{Top 10 expansion terms obtained by the word2vec and the relevance-based word embedding models for two sample queries  ``indian american museum'' and ``tibet protesters''.}
\begin{tabular}{|p{1.85cm}|p{1.85cm}|p{1.85cm}|p{1.85cm}||p{1.85cm}|p{1.85cm}|p{1.85cm}|p{1.85cm}|p{1.85cm}|} \hline
 \multicolumn{4}{|c||}{query: ``indian american museum''} & \multicolumn{4}{c|}{query: ``tibet protesters''}  \\\hline
\multicolumn{2}{|c|}{word2vec} & \multicolumn{2}{c||}{Rel.-based Embedding} & \multicolumn{2}{c|}{word2vec} & \multicolumn{2}{c|}{Rel.-based Embedding} \\\hline
\multicolumn{1}{|c|}{external} & \multicolumn{1}{c|}{target} & \multicolumn{1}{c|}{\firstmodel} & \multicolumn{1}{c||}{\secondmodel} & \multicolumn{1}{c|}{external} & \multicolumn{1}{c|}{target} & \multicolumn{1}{c|}{\firstmodel} & \multicolumn{1}{c|}{\secondmodel} \\\hline
 history & powwows & chumash & heye & demonstrators & tibetan & tibetan & tibetan \\
 art & smithsonian & heye & collection & protestors & lhasa & lama & tibetans \\
 culture & afro & artifacts & chumash & tibetan & demonstrators & tibetans & lama \\
 british & mesoamerica & smithsonian & smithsonian & protests & tibetans & lhasa & independence \\
 heritage & smithsonians & collection & york & tibetans & marchers & dalai & lhasa \\
 society & native & washington & new & protest & lhasas & independence & dalai \\
 states & heye & institution & apa & activists & jokhang & protest & open \\
 contemporary & hopi & york & native & protesting & demonstrations & open & protest \\
 part & mayas & native & americans & lhasa & dissidents & zone & zone \\
 united & cimam & apa & history & demonstrations & barkhor & followers & jokhang \\\hline

\end{tabular}
\label{tab:example}
\vspace{-0.2cm}
\end{table*}

\subsubsection{Results and Discussion}
To evaluate our models, we consider the following baselines: \textit{(i)} the standard maximum likelihood estimation (MLE) of the query model without query expansion, \textit{(ii)} two sets of embedding vectors (one trained on Google News as a large external corpus and one trained on the target retrieval collection) learned by the word2vec model\footnote{We use the CBOW implementation of the word2vec model. The skip-gram model also performs similarly.} \cite{Mikolov:2013}, and \textit{(iii)} two sets of embedding vectors (one trained on Wikipedia 2004 plus Gigawords 5 as a large external corpus\footnote{Available at \url{http://nlp.stanford.edu/projects/glove/}.} and the other on the target retrieval collection) learned by the GloVe model \cite{Pennington:2014}.

\tablename~\ref{tab:main} reports the results achieved by the proposed models and the baselines. According to this table, all the query expansion models outperform the MLE baseline in nearly all cases, which indicates the effectiveness of employing high-dimensional word representations for query expansion. Similar observations have been made in \cite{Diaz:2016,Kuzi:2016,Zamani:2016:ICTIR:emb,Zamani:2016:ICTIR:pqv}. According to the results, although word2vec performs slightly better than GloVe, no significant differences can be observed between their performances. According to \tablename~\ref{tab:main}, both relevance-based embedding models outperform all the baselines in all the collections, which shows the importance of taking relevance into account for training embedding vectors. These improvements are often statistically significant compared to all the baselines. The relevance likelihood maximization model (RLM) performs better than the relevance posterior estimation model (RPE) in all cases and the reason is related to their objective function. \firstmodel learns the relevance distribution for all terms, while \secondmodel learns the classification probability of being relevance for vocabulary terms (see Equations~\eqref{eq:model1:prob} and \eqref{eq:model2:prob}).

To get a sense of what is learned by each of the embedding models\footnote{For the sake of space, we only report the expanded terms estimated by the word2vec model and the proposed models.}, in \tablename~\ref{tab:example} we report the top 10 expansion terms for two sample queries from the Robust collection. According to this table, the terms added to the query by the word2vec model are syntactically or semantically related to individual query terms, which is expected. For the query ``indian american museum'' as an example, the terms ``history'', ``art'', and ``culture'' are related to the query term ``museum'', while the terms ``united'' and ``states'' are related to the query term ``american''. In contrast, looking at the expansion terms obtained by the relevance-based word embeddings, we can see that some relevant terms to the whole query were selected. For instance, ``chumash'' (a group of native americans)\footnote{see \url{https://en.wikipedia.org/wiki/Chumash_people}}, ``heye'' (the national museum of the American Indian in New York), ``smithsonian'' (the national museum of the American Indian in Washington DC), and ``apa'' (the American Psychological Association that actively promotes American Indian museums). A similar observation can be made for the other sample query (i.e., ``tibet protesters''). For example, the word ``independence'' is related to the whole query that was only selected by the relevance-based word embedding models, while the terms ``protestors'', ``protests'', ``protest'', and ``protesting'' that are syntactically similar to the query term ``protesters'' were considered by the word2vec model. We believe that these differences are due to the learning objective of the models. Interestingly, the expansion terms added to each query by the two relevance-based models look very similar, but according to \tablename~\ref{tab:main}, their performances are quite different. The reason is related to the weights given to each term by the two models. The weights given to the expansion terms by \secondmodel are very close to each other because its objective is to just classify each term and all of these terms are classified with a high probability as ``relevant''.

\begin{table}[t]
\centering
\caption{Evaluating relevance-based word embedding in pseudo-relevance feedback scenario. The superscripts 1/2/3 denote that the MAP improvements over RM3/Local Embedding/ERM with Local Embedding are statistically significant. The highest value in each row is marked in bold.}
\vspace{-0.2cm}
\setlength\tabcolsep{4pt}
\begin{tabular}{|c|l|c|c|c|l|} \hline
\multirow{2}{*}{Collection} & \multicolumn{1}{c|}{\multirow{2}{*}{Metric}} & \multirow{2}{*}{RM3} & Local & \multicolumn{2}{c|}{ERM} \topspace 
\\ \cline{5-6}
 &  &  & Emb.  & \multicolumn{1}{c|}{Local} & \multicolumn{1}{c|}{RLM} \topspace 
\\ \hline

\multirow{3}{*}{AP}
& MAP     & 0.2927 & 0.2412 & 0.3047 & \textbf{0.3119}\textsuperscript{12} \topspace\\
& P@20    & 0.4034 & 0.3742 & 0.4105 & \textbf{0.4233}\textsuperscript{12} \\
& NDCG@20 & 0.4368 & 0.4173 & 0.4411 & \textbf{0.4495}\textsuperscript{123} \\
\hline

\multirow{3}{*}{Robust}
& MAP     & 0.2593 & 0.2235 & 0.2643 & \textbf{0.2761}\textsuperscript{123} \topspace\\
& P@20    & 0.3486 & 0.3366 & 0.3498 & \textbf{0.3605}\textsuperscript{123} \\
& NDCG@20 & 0.4011 & 0.3868 & 0.4080 & \textbf{0.4173}\textsuperscript{123} \\
\hline

\multirow{3}{*}{GOV2}
& MAP     & 0.2863 & 0.2748 & 0.2924 & \textbf{0.2986}\textsuperscript{123} \topspace\\
& P@20    & 0.5318 & 0.5271 & 0.5379 & \textbf{0.5417}\textsuperscript{12} \\
& NDCG@20 & 0.4503 & 0.4576 & 0.4584 & \textbf{0.4603}\textsuperscript{123} \\
\hline

\multirow{3}{*}{ClueWeb}
& MAP     & 0.1079 & 0.1041 & 0.1094 & \textbf{0.1121}\textsuperscript{12} \topspace\\
& P@20    & 0.3111 & 0.3062 & 0.3145 & \textbf{0.3168} \\
& NDCG@20 & 0.2309 & 0.2261 & 0.2328 & \textbf{0.2360}\textsuperscript{2} \\
\hline

\end{tabular}
\label{tab:fb}
\vspace{-0.2cm}
\end{table}

\begin{figure*}[t]
\centering
\vspace{-0.35cm}
\begin{minipage}{.32\linewidth}
\centering
\subfloat[\# expansion terms]{\includegraphics[scale=.3]{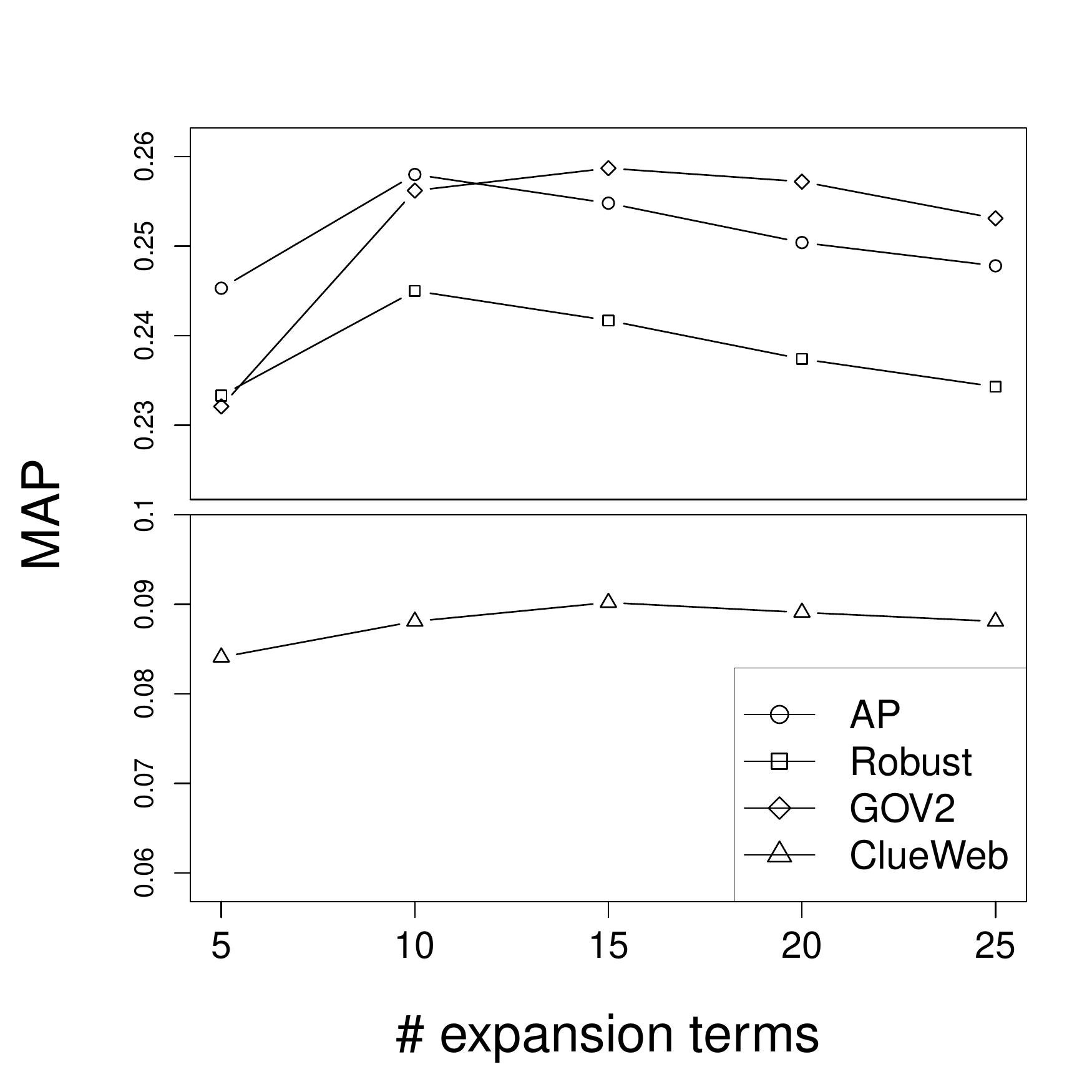}\label{fig:param_sens:a}}
\end{minipage}%
\begin{minipage}{.32\linewidth}
\centering
\subfloat[interpolation coefficient]{\includegraphics[scale=.3]{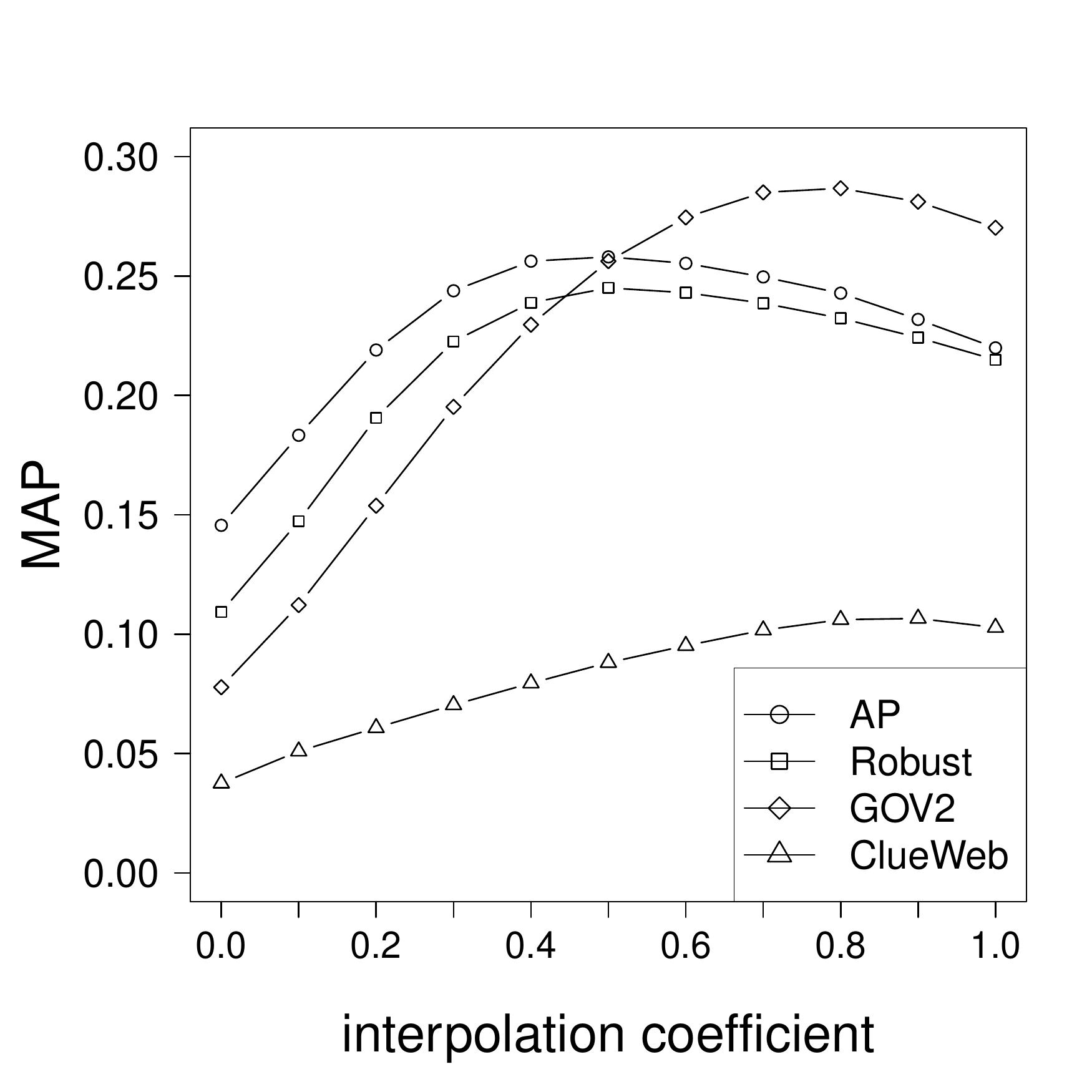}\label{fig:param_sens:b}}
\end{minipage}
\vspace{-0.2cm}
\caption{Sensitivity of \firstmodel to the number of expansion terms and the interpolation coefficient ($\alpha$), in terms of MAP.}
\label{fig:param_sens} 
\end{figure*}

In the next set of experiments, we consider the methods that use the top retrieved documents for query expansion: the relevance model (RM3) \cite{Abdul-jaleel:2004,Lavrenko:2001} as a state-of-the-art pseudo-relevance feedback model, and the local embedding approach recently proposed by Diaz et al. \cite{Diaz:2016} with the general idea of training word embedding models on the top ranked documents retrieved in response to a given query. Similar to \cite{Diaz:2016}, we use the word2vec model to train word embedding vectors on top $1000$ documents. The results are reported in \tablename~\ref{tab:fb}. In this table, ERM refers to the embedding-based relevance model recently proposed by Zamani and Croft \cite{Zamani:2016:ICTIR:emb} in order to make use of semantic similarities estimated based on the word embedding vectors in a pseudo-relevance feedback scenario. According to \tablename~\ref{tab:fb}, the ERM model that uses the relevance-based word embedding (RLM\footnote{For the sake of space, we only consider \firstmodel which shows better performance compared to \secondmodel in query expansion.}) outperforms all the other methods. These improvements are statistically significant in most cases. By comparing the results obtained by local embedding and those reported in \tablename~\ref{tab:main}, it can be observed that there are no substantial differences between the results for local embedding and word2vec. This is similar to what is reported by Diaz et al. \cite{Diaz:2016} when the embedding vectors are trained on the top documents in the target collection, similar to our setting. Note that the relevance-based model was also trained on the target collection. 

An interesting observation from Tables \ref{tab:main} and \ref{tab:fb} is that the \firstmodel performance (without using pseudo-relevant documents) in Robust and GOV2 is very close to the RM3 performance, and is slightly better in the GOV2 collection. Note that RM3 needs two retrieval runs\footnote{Diaz \cite{Diaz:2015} showed that for precision-oriented tasks, the second retrieval run can be restricted to the initial rank list for improving the efficiency of PRF models. However, for recall-oriented metrics, e.g., MAP, the second retrieval helps a lot.} and uses top retrieved documents, while \firstmodel only needs one retrieval run. This is an important issue in many real-world applications, since the efficiency constraints do not always allow them to have two retrieval runs per query.

\begin{figure}[t]
\vspace{-0.7cm}
    \centering
    \includegraphics[scale=.3]{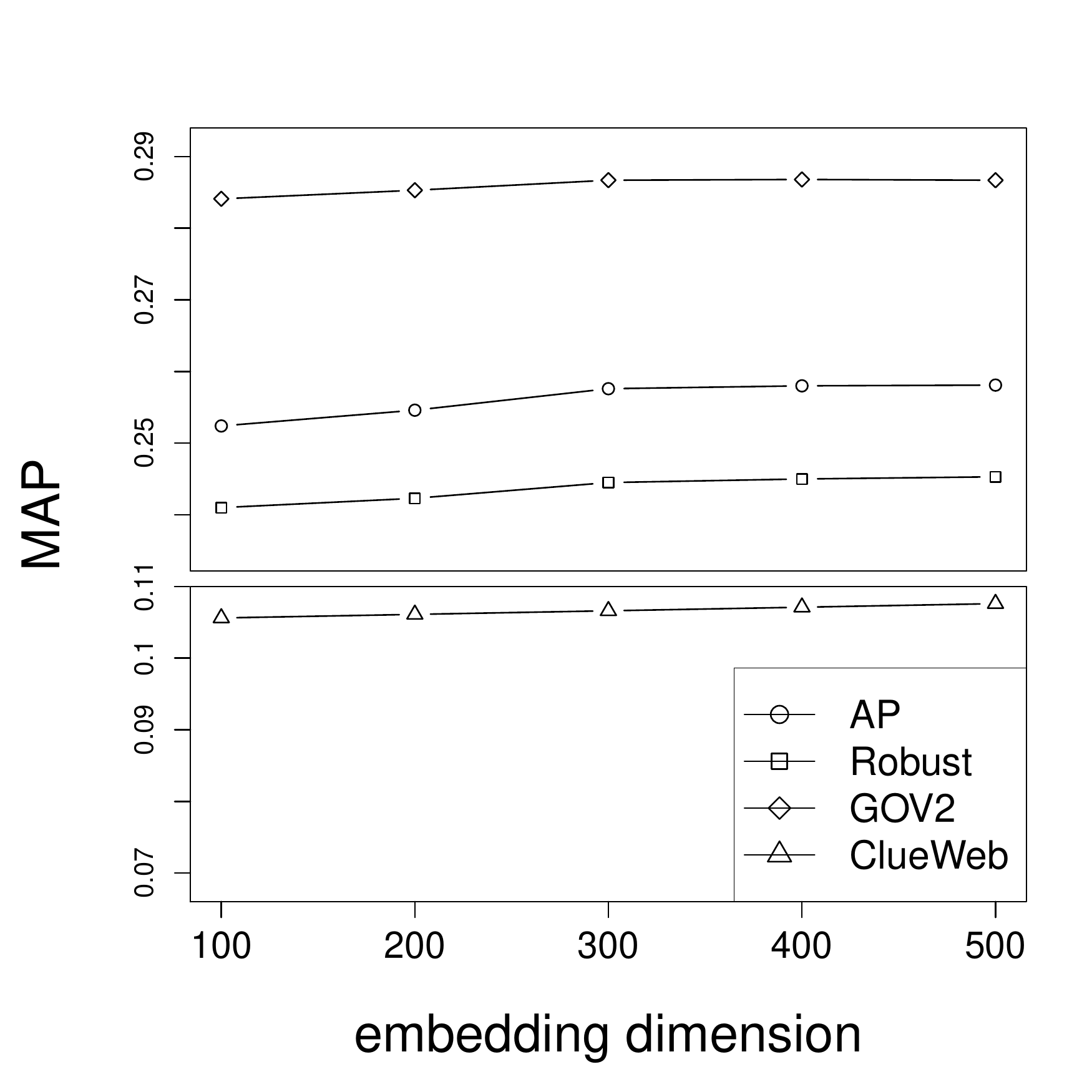}
    \vspace{-0.2cm}
    \caption{Sensitivity of \firstmodel to the dimension of embedding vectors, in terms of MAP.}
    \label{fig:dim}
    \vspace{-0.3cm}
\end{figure}

\textbf{Parameter Sensitivity.} In the next set of experiments, we study the sensitivity of \firstmodel as the best performing word embedding model in \tablename~\ref{tab:main} to the expansion parameters. \figurename~\ref{fig:param_sens:a} plots the sensitivity of \firstmodel to the number of expansion terms where the parameter $\alpha$ is set to $0.5$. According to this figure, in both newswire collections, the method shows its best performance when the queries are expanded with only $10$ words. In the GOV2 collection, $15$ words are needed for the method to show its best performance.

\figurename~\ref{fig:param_sens:b} plots the sensitivity of the methods to the interpolation coefficient $\alpha$ (see \eqname~\ref{eq:expansion}) where the number of expansion terms is set to $10$. According to the curves correspond to AP and Robust, the original query language model needs to be interpolated with the model estimated using relevance-based word embeddings with equal weights (i.e., $\alpha = 0.5$). This shows the quality of the estimated distribution via the learned embedding vectors. In the GOV2 collection, a higher weight should be given to the original query model, which indicates that the original query plays a key role in achieving good retrieval performance in this collection.

We also study the performance of \firstmodel as the best performing word embedding model for query expansion with respect to the embedding dimensionality. The results are shown in \figurename~\ref{fig:dim}, where the query expansion performance generally improves as we increase the embedding dimensionality. The performances become stable when the dimension is larger than $300$. This experiment suggests that $400$ dimensions would be enough for the relevance-based embedding model.

\begin{figure}[t]
\vspace{-0.7cm}
    \centering
    \includegraphics[scale=.3]{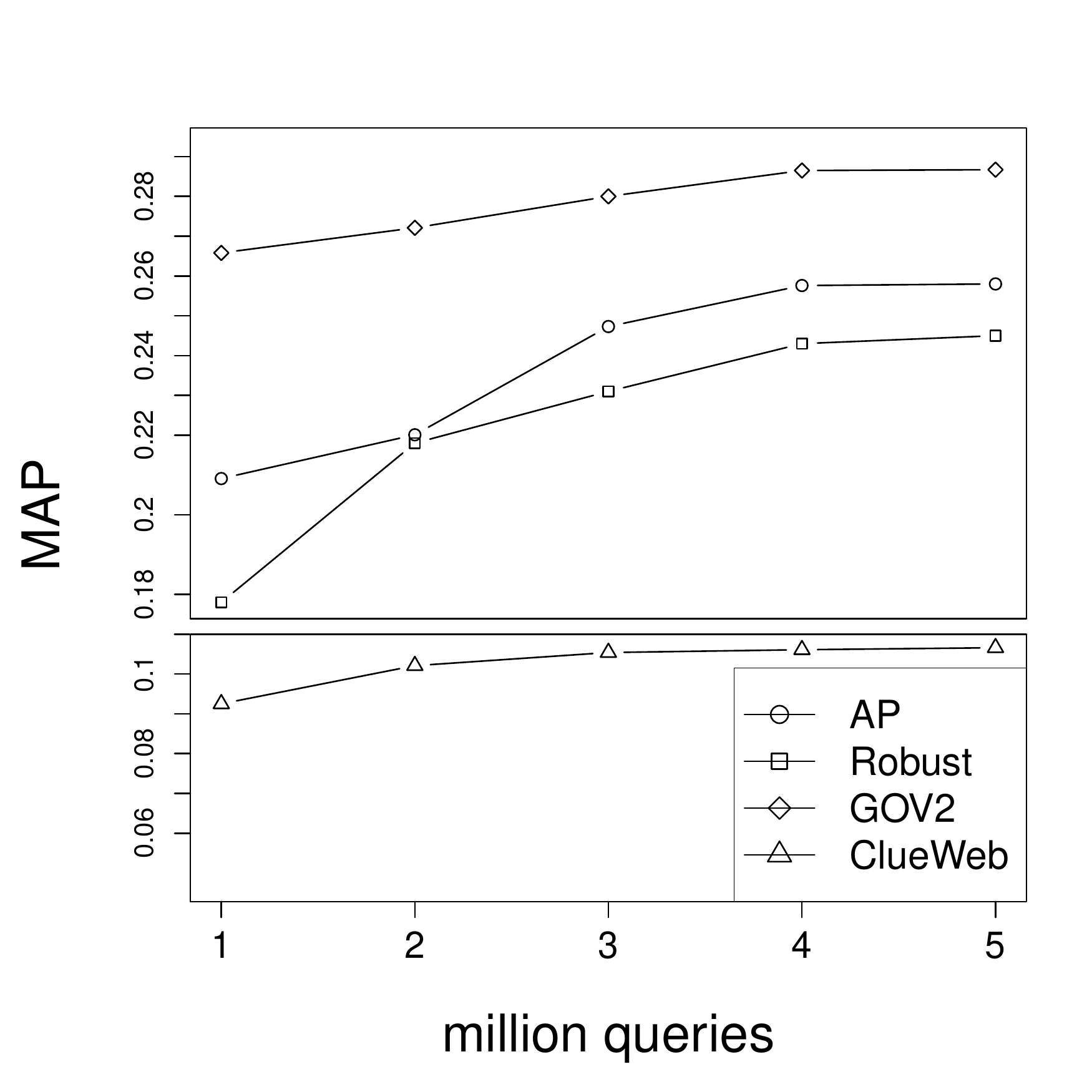}
    \vspace{-0.2cm}
    \caption{The Performance of \firstmodel with respect to different amount of training data (training queries), in terms of MAP.}
    \label{fig:lcurve}
    \vspace{-0.4cm}
\end{figure}

Due to the large number of parameters in the neural networks, they can require large amounts of training data to achieve good performance. In the next set of experiments, we study how much training data is needed for training our best model. The results are plotted in \figurename~\ref{fig:lcurve}. According to this figure, by increasing the number of training queries from one million to four million queries, the performance significantly increases, and becomes more stable after four million queries.

\subsection{Evaluation via Query Classification}
In this subsection, we evaluate the proposed embedding models in the context of query classification. In this task, each query is assigned to a number of labels (categories) which are pre-defined and a few training queries are available for each label. This is a supervised multi-label classification task with little training data.

\subsubsection{Data}
We consider the dataset that was introduced in KDD Cup 2005 \cite{Li:2005} for the internet user search query categorization task and was previously used in \cite{Zamani:2016:ICTIR:pqv} for evaluating query embedding vectors. This dataset contains $800$ web queries submitted by real users randomly collected from the MSN search logs. The queries do not contain ``junk'' text or non-English terms. The queries were labelled by three human editors. $67$ categories were pre-defined and up to $5$ labels were selected for each query by each editor.

\begin{table}[t]
    \centering
    \caption{Evaluating embedding algorithms via query classification. The superscripts 1/2 denote that the improvements over word2vec/GloVe are significant. The highest value in each column is marked in bold.}
    \vspace{-0.2cm}
    \begin{tabular}{|l|l|l|}\hline
        Method  & Precision & F1-measure  \\\hline
        word2vec & 0.3712 & 0.4008 \\\hline
        GloVe & 0.3643 & 0.3912 \\\hline
        Rel.-based Embedding - \firstmodel & 0.3943\textsuperscript{12} & 0.4267\textsuperscript{12} \\\hline
        Rel.-based Embedding - \secondmodel & \textbf{0.3961}\textsuperscript{12} & \textbf{0.4294}\textsuperscript{12} \\\hline
    \end{tabular}
    \label{tab:qc}
    \vspace{-0.3cm}
\end{table}

\subsubsection{Experimental Setup}
In our experiments, we performed 5-fold cross-validation over the queries and the reported results are the average of those obtained over the test folds. In all experiments, the spelling errors in queries were corrected in a pre-processing phase, the stopwords were removed from queries (using the INQUERY stopword list), and no stemming was performed.

To classify each query, we consider a very simple kNN-based approach proposed in \cite{Zamani:2016:ICTIR:pqv}. We first compute the probability of each category/label given each query $q$ and then select the top $t$ categories with the highest probabilities. The probability $p (C_i | q)$ is computed as follows:
\begin{equation}
    p (C_i | q) = \frac{\delta(\vec{C_i}, \vec{q})}{\sum_{j}{\delta(\vec{C_j}, \vec{q})}} \propto \delta(\vec{C_i}, \vec{q})
\end{equation}
where $C_i$ denotes the $i^{th}$ category. $\vec{C_i}$ is the centroid vector of all query embedding vectors with the label of $C_i$ in the training set. We ignore the query terms whose embedding vectors are not available. The number of labels assigned to each query was tuned on the training set from $\{1, 2, 3, 4, 5\}$. In the query classification experiments, we trained relevance-based word embedding using Robust as the collection. 

\subsubsection{Evaluation Metrics} We consider two evaluation metrics that were also used in KDD Cup 2005 \cite{Li:2005}: precision and F1-measure. Since the labels assigned by the three human editors differ in some cases, all the label sets should be taken into account. These metrics are computed in the same way as what is described in \cite{Li:2005} for evaluating the KDD Cup 2005 submitted runs. Statistically significant differences are determined using the two-tailed paired t-test computed at a $95\%$ confidence level ($p-value < 0.05$).

\subsubsection{Results and Discussion}
We compare our models against the word2vec and GloVe methods trained on the external collections that are described in the query expansion experiments. The results are reported in \tablename~\ref{tab:qc}, where the relevance-based embedding models significantly outperform the baselines in terms of both metrics. An interesting observation here is that contrary to the query expansion experiments, \secondmodel performs better than \firstmodel in query classification. The reason is that in query expansion the weight of each term is considered in order to generate the expanded query language model. Therefore, in addition to the order of terms, their weights should be also effective for improving the retrieval performance with query expansion. In query classification, we only assign a few categories to each query, and thus as long as the order of categories is correct, the similarity values between the queries and the categories do not matter. 

\begin{figure}[t]
    \centering
    \vspace{-0.5cm}
    \includegraphics[scale=.3]{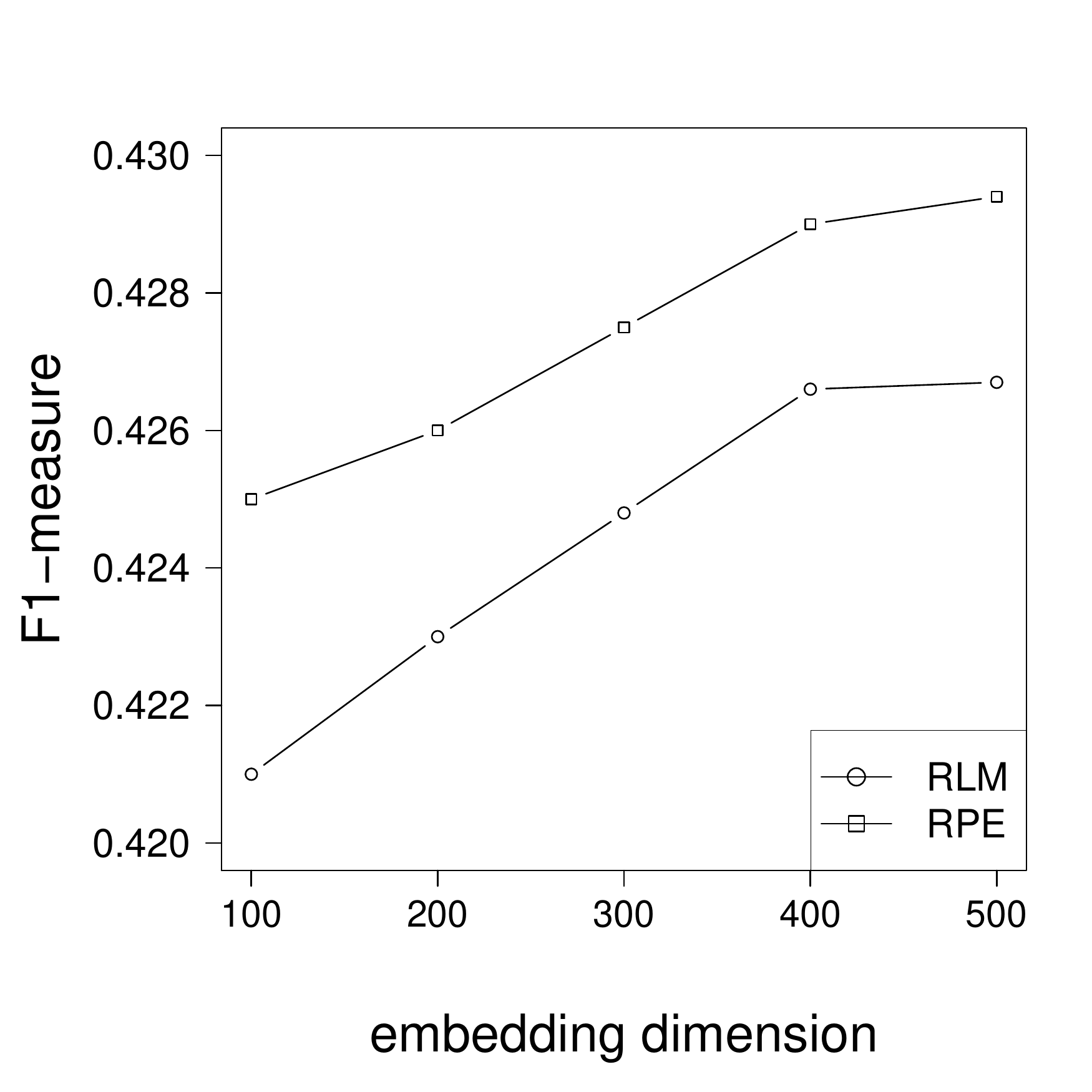}\vspace{-0.3cm}
    \caption{Sensitivity of the relevance-based embedding models to the embedding dimensionality, in terms of F1-measure.}
    \vspace{-0.4cm}
    \label{fig:dim-qc}
\end{figure}

\begin{figure}[t]
\vspace{-0.5cm}
    \centering
    \includegraphics[scale=.3]{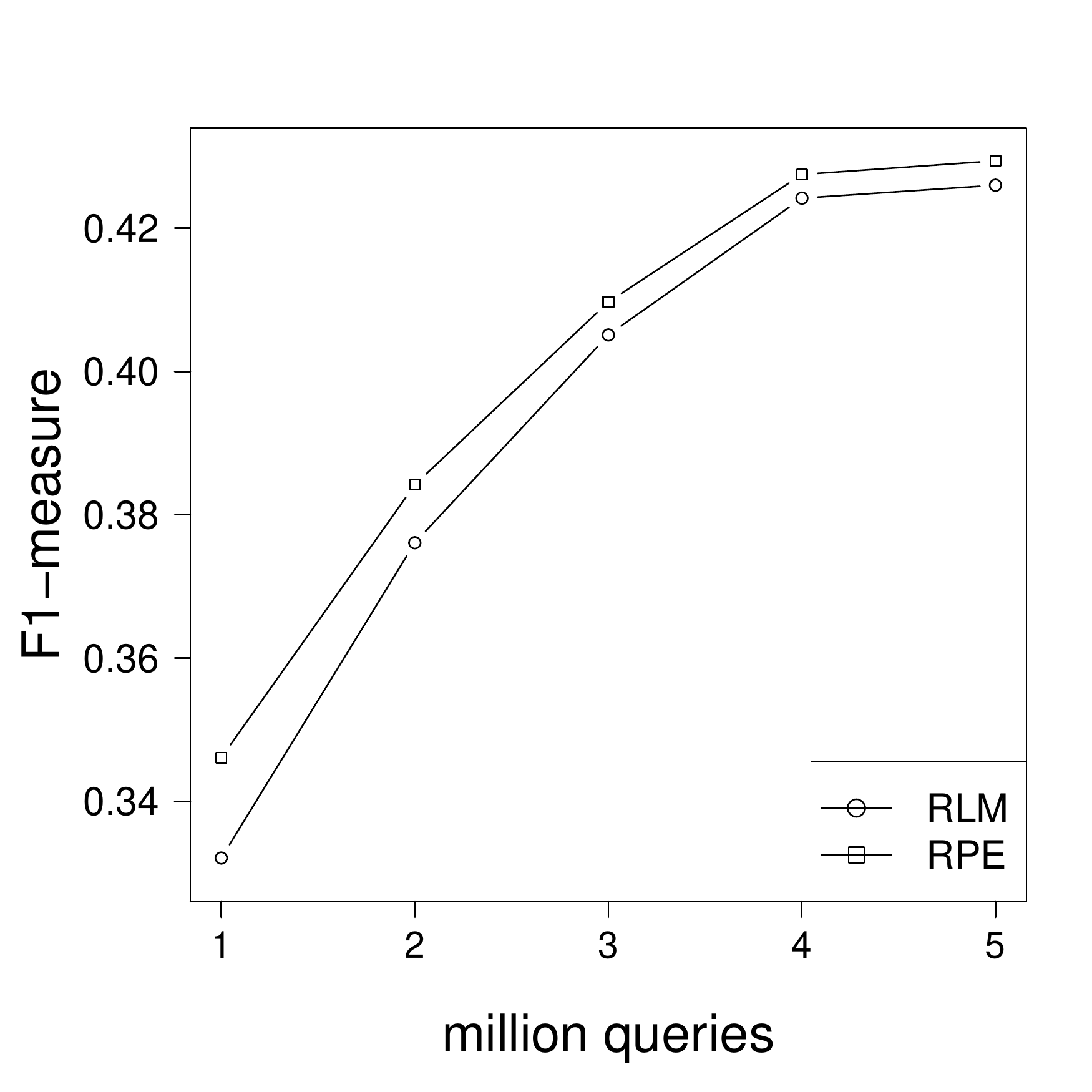}\vspace{-0.3cm}
    \caption{The Performance of relevance-based embedding models with respect to different amount of training data (training queries), in terms of F1-measure.}
    \vspace{-0.4cm}
    \label{fig:lcurve-qc}
\end{figure}

In the next set of experiments, we study the performance of our relevance-based word embedding models with respect to the embedding dimensionality. The results are plotted in \figurename~\ref{fig:dim-qc}. According to this figure, the performance is generally improved by increasing the embedding dimensionality, and becomes stable when the dimension is greater than $400$. This is similar to our observation in the query expansion experiments. We also study the amount of data needed for training our models in \figurename~\ref{fig:lcurve-qc}. According to this figure, at least $4$ million queries are needed in order to learn accurate relevance-based word embeddings. It can be seen from \figurename~\ref{fig:lcurve-qc} that \firstmodel needs more training data compared to \secondmodel in order to perform well, because by increasing the amount of training data the learning curves of these two models get closer.

\vspace{-0.1cm}
\section{Conclusions and Future Work}
\label{sec:conclusion}
In this paper, we revisited the underlying assumption in typical word embedding models, such as word2vec and GloVe. Instead of learning embedding vectors based on term proximity, we proposed learning embeddings based on the notion of relevance, which is the primary objective in many IR tasks. We developed two neural network-based models for learning relevance-based word embeddings. The first model, the relevance likelihood maximization model, aims to estimate the probability of each word in a relevance distribution for each query, while the second one, the relevance posterior estimation model, classifies each term as belonging to relevant or non-relevant class for each query. We evaluated our models using two sets of extrinsic evaluation: query expansion and query classification. The query expansion experiments using four standard TREC collections, two newswire and two large-scale web collections, suggested that the relevance-based word embedding models outperform state-of-the-art word embedding algorithms. We showed that the expansion terms chosen by our models are related to the whole query, while those chosen by typical word embedding models are related to individual query terms. The query classification experiments also validated these findings and investigated the effectiveness of our models.

In the future, we intend to evaluate the learned embedding models in other IR tasks, such as query reformulation, query intent prediction, etc. We can also achieve more accurate relevance-based embedding vectors by considering the clicked documents for training query, instead of or in addition to the top retrieved documents. 

\medskip
\small
\noindent {\textbf{\textit{Acknowledgements.}}}
The authors thank Daniel Cohen, Mostafa Dehghani, and Qingyao Ai for their invaluable comments. This work was supported in part by the Center for Intelligent Information Retrieval. Any opinions, findings and conclusions or recommendations expressed in this material are those of the authors and do not necessarily reflect those of the sponsor. 

\vspace{-0.1cm}


\begin{thebibliography}{00}


\ifx \showCODEN    \undefined \def \showCODEN     #1{\unskip}     \fi
\ifx \showDOI      \undefined \def \showDOI       #1{{\tt DOI:}\penalty0{#1}\ }
  \fi
\ifx \showISBNx    \undefined \def \showISBNx     #1{\unskip}     \fi
\ifx \showISBNxiii \undefined \def \showISBNxiii  #1{\unskip}     \fi
\ifx \showISSN     \undefined \def \showISSN      #1{\unskip}     \fi
\ifx \showLCCN     \undefined \def \showLCCN      #1{\unskip}     \fi
\ifx \shownote     \undefined \def \shownote      #1{#1}          \fi
\ifx \showarticletitle \undefined \def \showarticletitle #1{#1}   \fi
\ifx \showURL      \undefined \def \showURL       #1{#1}          \fi
\providecommand\bibfield[2]{#2}
\providecommand\bibinfo[2]{#2}
\providecommand\natexlab[1]{#1}
\providecommand\showeprint[2][]{arXiv:#2}

\bibitem[\protect\citeauthoryear{Abdul-jaleel, Allan, Croft, Diaz, Larkey, Li,
  Metzler, Smucker, Strohman, Turtle, and Wade}{Abdul-jaleel
  et~al\mbox{.}}{2004}]%
        {Abdul-jaleel:2004}
\bibfield{author}{\bibinfo{person}{Nasreen Abdul-jaleel},
  \bibinfo{person}{James Allan}, \bibinfo{person}{W.~Bruce Croft},
  \bibinfo{person}{Fernando Diaz}, \bibinfo{person}{Leah Larkey},
  \bibinfo{person}{Xiaoyan Li}, \bibinfo{person}{Donald Metzler},
  \bibinfo{person}{Mark~D. Smucker}, \bibinfo{person}{Trevor Strohman},
  \bibinfo{person}{Howard Turtle}, {and} \bibinfo{person}{Courtney Wade}.}
  \bibinfo{year}{2004}\natexlab{}.
\newblock \showarticletitle{UMass at TREC 2004: Novelty and HARD}. In
  \bibinfo{booktitle}{{\em TREC '04}}.
\newblock


\bibitem[\protect\citeauthoryear{Ai, Yang, Guo, and Croft}{Ai
  et~al\mbox{.}}{2016}]%
        {Ai:2016}
\bibfield{author}{\bibinfo{person}{Qingyao Ai}, \bibinfo{person}{Liu Yang},
  \bibinfo{person}{Jiafeng Guo}, {and} \bibinfo{person}{W.~Bruce Croft}.}
  \bibinfo{year}{2016}\natexlab{}.
\newblock \showarticletitle{Analysis of the Paragraph Vector Model for
  Information Retrieval}. In \bibinfo{booktitle}{{\em ICTIR '16}}.
  \bibinfo{pages}{133--142}.
\newblock


\bibitem[\protect\citeauthoryear{Amati and Van~Rijsbergen}{Amati and
  Van~Rijsbergen}{2002}]%
        {Amati:2002}
\bibfield{author}{\bibinfo{person}{Gianni Amati} {and}
  \bibinfo{person}{Cornelis~Joost Van~Rijsbergen}.}
  \bibinfo{year}{2002}\natexlab{}.
\newblock \showarticletitle{Probabilistic Models of Information Retrieval Based
  on Measuring the Divergence from Randomness}.
\newblock \bibinfo{journal}{{\em ACM Trans. Inf. Syst.\/}}
  \bibinfo{volume}{20}, \bibinfo{number}{4} (\bibinfo{year}{2002}),
  \bibinfo{pages}{357--389}.
\newblock


\bibitem[\protect\citeauthoryear{Bruza and Song}{Bruza and Song}{2002}]%
        {Bruza:2002}
\bibfield{author}{\bibinfo{person}{P.~D. Bruza} {and} \bibinfo{person}{D.
  Song}.} \bibinfo{year}{2002}\natexlab{}.
\newblock \showarticletitle{Inferring Query Models by Computing Information
  Flow}. In \bibinfo{booktitle}{{\em CIKM '02}}. \bibinfo{pages}{260--269}.
\newblock


\bibitem[\protect\citeauthoryear{Clinchant and Perronnin}{Clinchant and
  Perronnin}{2013}]%
        {Clinchant:2013}
\bibfield{author}{\bibinfo{person}{Stephane Clinchant} {and}
  \bibinfo{person}{Florent Perronnin}.} \bibinfo{year}{2013}\natexlab{}.
\newblock \showarticletitle{{Aggregating Continuous Word Embeddings for
  Information Retrieval}}. In \bibinfo{booktitle}{{\em CVSC@ACL '13}}.
  \bibinfo{pages}{100--109}.
\newblock


\bibitem[\protect\citeauthoryear{Cormack, Smucker, and Clarke}{Cormack
  et~al\mbox{.}}{2011}]%
        {Cormack:2011}
\bibfield{author}{\bibinfo{person}{Gordon~V. Cormack}, \bibinfo{person}{Mark~D.
  Smucker}, {and} \bibinfo{person}{Charles~L. Clarke}.}
  \bibinfo{year}{2011}\natexlab{}.
\newblock \showarticletitle{Efficient and Effective Spam Filtering and
  Re-ranking for Large Web Datasets}.
\newblock \bibinfo{journal}{{\em Inf. Retr.\/}} \bibinfo{volume}{14},
  \bibinfo{number}{5} (\bibinfo{year}{2011}), \bibinfo{pages}{441--465}.
\newblock


\bibitem[\protect\citeauthoryear{Croft, Metzler, and Strohman}{Croft
  et~al\mbox{.}}{2009}]%
        {Croft:2009}
\bibfield{author}{\bibinfo{person}{Bruce Croft}, \bibinfo{person}{Donald
  Metzler}, {and} \bibinfo{person}{Trevor Strohman}.}
  \bibinfo{year}{2009}\natexlab{}.
\newblock \bibinfo{booktitle}{{\em Search Engines: Information Retrieval in
  Practice\/} (\bibinfo{edition}{1st} ed.)}.
\newblock \bibinfo{publisher}{Addison-Wesley Publishing Company}.
\newblock
\showISBNx{0136072240, 9780136072249}


\bibitem[\protect\citeauthoryear{Deerwester, Dumais, Furnas, Landauer, and
  Harshman}{Deerwester et~al\mbox{.}}{1990}]%
        {Deerwester:1988}
\bibfield{author}{\bibinfo{person}{Scott Deerwester}, \bibinfo{person}{Susan~T.
  Dumais}, \bibinfo{person}{George~W. Furnas}, \bibinfo{person}{Thomas~K.
  Landauer}, {and} \bibinfo{person}{Richard Harshman}.}
  \bibinfo{year}{1990}\natexlab{}.
\newblock \showarticletitle{{Indexing by Latent Semantic Analysis}}.
\newblock  \bibinfo{volume}{41}, \bibinfo{number}{6} (\bibinfo{year}{1990}),
  \bibinfo{pages}{391--407}.
\newblock


\bibitem[\protect\citeauthoryear{Dehghani, Zamani, Severyn, Kamps, and
  Croft}{Dehghani et~al\mbox{.}}{2017}]%
        {Dehghani:2017}
\bibfield{author}{\bibinfo{person}{Mostafa Dehghani}, \bibinfo{person}{Hamed
  Zamani}, \bibinfo{person}{Aliaksei Severyn}, \bibinfo{person}{Jaap Kamps},
  {and} \bibinfo{person}{W.~Bruce Croft}.} \bibinfo{year}{2017}\natexlab{}.
\newblock \showarticletitle{Neural Ranking Models with Weak Supervision}. In
  \bibinfo{booktitle}{{\em SIGIR '17}}.
\newblock


\bibitem[\protect\citeauthoryear{Diaz}{Diaz}{2015}]%
        {Diaz:2015}
\bibfield{author}{\bibinfo{person}{Fernando Diaz}.}
  \bibinfo{year}{2015}\natexlab{}.
\newblock \showarticletitle{Condensed List Relevance Models}. In
  \bibinfo{booktitle}{{\em ICTIR '15}}. \bibinfo{pages}{313--316}.
\newblock


\bibitem[\protect\citeauthoryear{Diaz, Mitra, and Craswell}{Diaz
  et~al\mbox{.}}{2016}]%
        {Diaz:2016}
\bibfield{author}{\bibinfo{person}{Fernando Diaz}, \bibinfo{person}{Bhaskar
  Mitra}, {and} \bibinfo{person}{Nick Craswell}.}
  \bibinfo{year}{2016}\natexlab{}.
\newblock \showarticletitle{{Query Expansion with Locally-Trained Word
  Embeddings}}. In \bibinfo{booktitle}{{\em ACL '16}}.
\newblock


\bibitem[\protect\citeauthoryear{Gutmann and Hyv\"{a}rinen}{Gutmann and
  Hyv\"{a}rinen}{2012}]%
        {Gutmann:2012}
\bibfield{author}{\bibinfo{person}{Michael~U. Gutmann} {and}
  \bibinfo{person}{Aapo Hyv\"{a}rinen}.} \bibinfo{year}{2012}\natexlab{}.
\newblock \showarticletitle{Noise-contrastive Estimation of Unnormalized
  Statistical Models, with Applications to Natural Image Statistics}.
\newblock \bibinfo{journal}{{\em J. Mach. Learn. Res.\/}} \bibinfo{volume}{13},
  \bibinfo{number}{1} (\bibinfo{year}{2012}), \bibinfo{pages}{307--361}.
\newblock


\bibitem[\protect\citeauthoryear{J\"{a}rvelin and
  Kek\"{a}l\"{a}inen}{J\"{a}rvelin and Kek\"{a}l\"{a}inen}{2002}]%
        {Jarvelin:2002}
\bibfield{author}{\bibinfo{person}{Kalervo J\"{a}rvelin} {and}
  \bibinfo{person}{Jaana Kek\"{a}l\"{a}inen}.} \bibinfo{year}{2002}\natexlab{}.
\newblock \showarticletitle{Cumulated Gain-based Evaluation of IR Techniques}.
\newblock \bibinfo{journal}{{\em ACM Trans. Inf. Syst.\/}}
  \bibinfo{volume}{20}, \bibinfo{number}{4} (\bibinfo{date}{Oct.}
  \bibinfo{year}{2002}), \bibinfo{pages}{422--446}.
\newblock


\bibitem[\protect\citeauthoryear{Jing and Croft}{Jing and Croft}{1994}]%
        {Jing:1994}
\bibfield{author}{\bibinfo{person}{Yufeng Jing} {and} \bibinfo{person}{W.~Bruce
  Croft}.} \bibinfo{year}{1994}\natexlab{}.
\newblock \showarticletitle{An Association Thesaurus for Information
  Retrieval}. In \bibinfo{booktitle}{{\em RIAO '94}}.
  \bibinfo{pages}{146--160}.
\newblock


\bibitem[\protect\citeauthoryear{Kenter and de~Rijke}{Kenter and
  de~Rijke}{2015}]%
        {Kenter:2015}
\bibfield{author}{\bibinfo{person}{Tom Kenter} {and} \bibinfo{person}{Maarten
  de Rijke}.} \bibinfo{year}{2015}\natexlab{}.
\newblock \showarticletitle{{Short Text Similarity with Word Embeddings}}. In
  \bibinfo{booktitle}{{\em CIKM '15}}. \bibinfo{pages}{1411--1420}.
\newblock


\bibitem[\protect\citeauthoryear{Kusner, Sun, Kolkin, and Weinberger}{Kusner
  et~al\mbox{.}}{2015}]%
        {Kusner:2015}
\bibfield{author}{\bibinfo{person}{Matt~J. Kusner}, \bibinfo{person}{Yu Sun},
  \bibinfo{person}{Nicholas~I. Kolkin}, {and} \bibinfo{person}{Kilian~Q.
  Weinberger}.} \bibinfo{year}{2015}\natexlab{}.
\newblock \showarticletitle{{From Word Embeddings to Document Distances}}. In
  \bibinfo{booktitle}{{\em ICML '15}}. \bibinfo{pages}{957--966}.
\newblock


\bibitem[\protect\citeauthoryear{Kuzi, Shtok, and Kurland}{Kuzi
  et~al\mbox{.}}{2016}]%
        {Kuzi:2016}
\bibfield{author}{\bibinfo{person}{Saar Kuzi}, \bibinfo{person}{Anna Shtok},
  {and} \bibinfo{person}{Oren Kurland}.} \bibinfo{year}{2016}\natexlab{}.
\newblock \showarticletitle{Query Expansion Using Word Embeddings}. In
  \bibinfo{booktitle}{{\em CIKM '16}}. \bibinfo{pages}{1929--1932}.
\newblock


\bibitem[\protect\citeauthoryear{Lafferty and Zhai}{Lafferty and Zhai}{2001}]%
        {Lafferty:2001}
\bibfield{author}{\bibinfo{person}{John Lafferty} {and}
  \bibinfo{person}{Chengxiang Zhai}.} \bibinfo{year}{2001}\natexlab{}.
\newblock \showarticletitle{{Document Language Models, Query Models, and Risk
  Minimization for Information Retrieval}}. In \bibinfo{booktitle}{{\em SIGIR
  '01}}. \bibinfo{pages}{111--119}.
\newblock


\bibitem[\protect\citeauthoryear{Lavrenko, Choquette, and Croft}{Lavrenko
  et~al\mbox{.}}{2002}]%
        {Lavrenko:2002}
\bibfield{author}{\bibinfo{person}{Victor Lavrenko}, \bibinfo{person}{Martin
  Choquette}, {and} \bibinfo{person}{W.~Bruce Croft}.}
  \bibinfo{year}{2002}\natexlab{}.
\newblock \showarticletitle{Cross-lingual Relevance Models}. In
  \bibinfo{booktitle}{{\em SIGIR '02}}. \bibinfo{pages}{175--182}.
\newblock


\bibitem[\protect\citeauthoryear{Lavrenko and Croft}{Lavrenko and
  Croft}{2001}]%
        {Lavrenko:2001}
\bibfield{author}{\bibinfo{person}{Victor Lavrenko} {and}
  \bibinfo{person}{W.~Bruce Croft}.} \bibinfo{year}{2001}\natexlab{}.
\newblock \showarticletitle{{Relevance Based Language Models}}. In
  \bibinfo{booktitle}{{\em SIGIR '01}}. \bibinfo{pages}{120--127}.
\newblock


\bibitem[\protect\citeauthoryear{Levy and Goldberg}{Levy and Goldberg}{2014}]%
        {Levy:2014}
\bibfield{author}{\bibinfo{person}{Omer Levy} {and} \bibinfo{person}{Yoav
  Goldberg}.} \bibinfo{year}{2014}\natexlab{}.
\newblock \showarticletitle{Neural Word Embedding as Implicit Matrix
  Factorization}.
\newblock In \bibinfo{booktitle}{{\em NIPS '14}}. \bibinfo{pages}{2177--2185}.
\newblock


\bibitem[\protect\citeauthoryear{Li, Zheng, and Dai}{Li et~al\mbox{.}}{2005}]%
        {Li:2005}
\bibfield{author}{\bibinfo{person}{Ying Li}, \bibinfo{person}{Zijian Zheng},
  {and} \bibinfo{person}{Honghua~(Kathy) Dai}.}
  \bibinfo{year}{2005}\natexlab{}.
\newblock \showarticletitle{KDD CUP-2005 Report: Facing a Great Challenge}.
\newblock \bibinfo{journal}{{\em SIGKDD Explor. Newsl.\/}} \bibinfo{volume}{7},
  \bibinfo{number}{2} (\bibinfo{year}{2005}), \bibinfo{pages}{91--99}.
\newblock


\bibitem[\protect\citeauthoryear{Liu, Gao, He, Deng, Duh, and Wang}{Liu
  et~al\mbox{.}}{2015}]%
        {Liu:2015}
\bibfield{author}{\bibinfo{person}{Xiaodong Liu}, \bibinfo{person}{Jianfeng
  Gao}, \bibinfo{person}{Xiaodong He}, \bibinfo{person}{Li Deng},
  \bibinfo{person}{Kevin Duh}, {and} \bibinfo{person}{Ye-yi Wang}.}
  \bibinfo{year}{2015}\natexlab{}.
\newblock \showarticletitle{{Representation Learning Using Multi-Task Deep
  Neural Networks for Semantic Classification and Information Retrieval}}. In
  \bibinfo{booktitle}{{\em NAACL '15}}. \bibinfo{pages}{912--921}.
\newblock


\bibitem[\protect\citeauthoryear{Mikolov, Sutskever, Chen, Corrado, and
  Dean}{Mikolov et~al\mbox{.}}{2013}]%
        {Mikolov:2013}
\bibfield{author}{\bibinfo{person}{Tomas Mikolov}, \bibinfo{person}{Ilya
  Sutskever}, \bibinfo{person}{Kai Chen}, \bibinfo{person}{Greg~S Corrado},
  {and} \bibinfo{person}{Jeff Dean}.} \bibinfo{year}{2013}\natexlab{}.
\newblock \showarticletitle{{Distributed Representations of Words and Phrases
  and their Compositionality}}. In \bibinfo{booktitle}{{\em NIPS '13}}.
  \bibinfo{pages}{3111--3119}.
\newblock


\bibitem[\protect\citeauthoryear{Mnih and Hinton}{Mnih and Hinton}{2009}]%
        {Minh:2009}
\bibfield{author}{\bibinfo{person}{Andriy Mnih} {and}
  \bibinfo{person}{Geoffrey~E Hinton}.} \bibinfo{year}{2009}\natexlab{}.
\newblock \showarticletitle{A Scalable Hierarchical Distributed Language
  Model}.
\newblock In \bibinfo{booktitle}{{\em NIPS '09}}. \bibinfo{pages}{1081--1088}.
\newblock


\bibitem[\protect\citeauthoryear{Morin and Bengio}{Morin and Bengio}{2005}]%
        {Morin:2005}
\bibfield{author}{\bibinfo{person}{Frederic Morin} {and}
  \bibinfo{person}{Yoshua Bengio}.} \bibinfo{year}{2005}\natexlab{}.
\newblock \showarticletitle{Hierarchical Probabilistic Neural Network Language
  Model}. In \bibinfo{booktitle}{{\em AISTATS '05}}. \bibinfo{pages}{246--252}.
\newblock


\bibitem[\protect\citeauthoryear{Pass, Chowdhury, and Torgeson}{Pass
  et~al\mbox{.}}{2006}]%
        {Pass:2006}
\bibfield{author}{\bibinfo{person}{Greg Pass}, \bibinfo{person}{Abdur
  Chowdhury}, {and} \bibinfo{person}{Cayley Torgeson}.}
  \bibinfo{year}{2006}\natexlab{}.
\newblock \showarticletitle{A Picture of Search}. In \bibinfo{booktitle}{{\em
  InfoScale '06}}.
\newblock


\bibitem[\protect\citeauthoryear{Pennington, Socher, and Manning}{Pennington
  et~al\mbox{.}}{2014}]%
        {Pennington:2014}
\bibfield{author}{\bibinfo{person}{Jeffrey Pennington},
  \bibinfo{person}{Richard Socher}, {and} \bibinfo{person}{Christopher
  Manning}.} \bibinfo{year}{2014}\natexlab{}.
\newblock \showarticletitle{{GloVe: Global Vectors for Word Representation}}.
  In \bibinfo{booktitle}{{\em EMNLP '14}}. \bibinfo{pages}{1532--1543}.
\newblock


\bibitem[\protect\citeauthoryear{Ponte and Croft}{Ponte and Croft}{1998}]%
        {Ponte:1998}
\bibfield{author}{\bibinfo{person}{Jay~M. Ponte} {and}
  \bibinfo{person}{W.~Bruce Croft}.} \bibinfo{year}{1998}\natexlab{}.
\newblock \showarticletitle{{A Language Modeling Approach to Information
  Retrieval}}. In \bibinfo{booktitle}{{\em SIGIR '98}}.
  \bibinfo{pages}{275--281}.
\newblock


\bibitem[\protect\citeauthoryear{Rekabsaz, Lupu, Hanbury, and Zamani}{Rekabsaz
  et~al\mbox{.}}{2017}]%
        {Rekabsaz:2017}
\bibfield{author}{\bibinfo{person}{Navid Rekabsaz}, \bibinfo{person}{Mihai
  Lupu}, \bibinfo{person}{Allan Hanbury}, {and} \bibinfo{person}{Hamed
  Zamani}.} \bibinfo{year}{2017}\natexlab{}.
\newblock \showarticletitle{Word Embedding Causes Topic Shifting; Exploit
  Global Context!}. In \bibinfo{booktitle}{{\em SIGIR '17}}.
\newblock


\bibitem[\protect\citeauthoryear{Rekabsaz, Lupu, Hanbury, and Zuccon}{Rekabsaz
  et~al\mbox{.}}{2016}]%
        {Rekabsaz:2016}
\bibfield{author}{\bibinfo{person}{Navid Rekabsaz}, \bibinfo{person}{Mihai
  Lupu}, \bibinfo{person}{Allan Hanbury}, {and} \bibinfo{person}{Guido
  Zuccon}.} \bibinfo{year}{2016}\natexlab{}.
\newblock \showarticletitle{Generalizing Translation Models in the
  Probabilistic Relevance Framework}. In \bibinfo{booktitle}{{\em CIKM '16}}.
  \bibinfo{pages}{711--720}.
\newblock


\bibitem[\protect\citeauthoryear{Rocchio}{Rocchio}{1971}]%
        {Rocchio:1971}
\bibfield{author}{\bibinfo{person}{J.~J. Rocchio}.}
  \bibinfo{year}{1971}\natexlab{}.
\newblock \showarticletitle{{Relevance Feedback in Information Retrieval}}.
\newblock In \bibinfo{booktitle}{{\em The {SMART} Retrieval System: Experiments
  in Automatic Document Processing}}. \bibinfo{pages}{313--323}.
\newblock


\bibitem[\protect\citeauthoryear{Rumelhart, Hinton, and Williams}{Rumelhart
  et~al\mbox{.}}{1986}]%
        {Rumelhart:1986}
\bibfield{author}{\bibinfo{person}{D.~E. Rumelhart}, \bibinfo{person}{G.~E.
  Hinton}, {and} \bibinfo{person}{R.~J. Williams}.}
  \bibinfo{year}{1986}\natexlab{}.
\newblock \showarticletitle{Learning representations by back-propagating
  errors}.
\newblock \bibinfo{journal}{{\em Nature\/}}  \bibinfo{volume}{323}
  (\bibinfo{date}{Oct.} \bibinfo{year}{1986}), \bibinfo{pages}{533--536}.
\newblock


\bibitem[\protect\citeauthoryear{Saracevic}{Saracevic}{2016}]%
        {Saracevic:2016}
\bibfield{author}{\bibinfo{person}{T. Saracevic}.}
  \bibinfo{year}{2016}\natexlab{}.
\newblock \bibinfo{booktitle}{{\em The Notion of Relevance in Information
  Science: Everybody knows what relevance is. But, what is it really?}}
\newblock \bibinfo{publisher}{Morgan \& Claypool Publishers}.
\newblock
\showISBNx{9781598297690}


\bibitem[\protect\citeauthoryear{Sordoni, Bengio, and Nie}{Sordoni
  et~al\mbox{.}}{2014}]%
        {Sordoni:2014}
\bibfield{author}{\bibinfo{person}{Alessandro Sordoni}, \bibinfo{person}{Yoshua
  Bengio}, {and} \bibinfo{person}{Jian-Yun Nie}.}
  \bibinfo{year}{2014}\natexlab{}.
\newblock \showarticletitle{{Learning Concept Embeddings for Query Expansion by
  Quantum Entropy Minimization}}. In \bibinfo{booktitle}{{\em AAAI '14}}.
  \bibinfo{pages}{1586--1592}.
\newblock


\bibitem[\protect\citeauthoryear{Tao and Zhai}{Tao and Zhai}{2006}]%
        {Tao:2006b}
\bibfield{author}{\bibinfo{person}{Tao Tao} {and} \bibinfo{person}{ChengXiang
  Zhai}.} \bibinfo{year}{2006}\natexlab{}.
\newblock \showarticletitle{{Regularized Estimation of Mixture Models for
  Robust Pseudo-relevance Feedback}}. In \bibinfo{booktitle}{{\em SIGIR '06}}.
  \bibinfo{pages}{162--169}.
\newblock


\bibitem[\protect\citeauthoryear{Vuli\'{c} and Moens}{Vuli\'{c} and
  Moens}{2015}]%
        {Vulic:2015}
\bibfield{author}{\bibinfo{person}{Ivan Vuli\'{c}} {and}
  \bibinfo{person}{Marie-Francine Moens}.} \bibinfo{year}{2015}\natexlab{}.
\newblock \showarticletitle{{Monolingual and Cross-Lingual Information
  Retrieval Models Based on (Bilingual) Word Embeddings}}. In
  \bibinfo{booktitle}{{\em SIGIR '15}}. \bibinfo{pages}{363--372}.
\newblock


\bibitem[\protect\citeauthoryear{Xu and Croft}{Xu and Croft}{1996}]%
        {Xu:1996}
\bibfield{author}{\bibinfo{person}{Jinxi Xu} {and} \bibinfo{person}{W.~Bruce
  Croft}.} \bibinfo{year}{1996}\natexlab{}.
\newblock \showarticletitle{{Query Expansion Using Local and Global Document
  Analysis}}. In \bibinfo{booktitle}{{\em SIGIR '96}}. \bibinfo{pages}{4--11}.
\newblock


\bibitem[\protect\citeauthoryear{Zamani, Bendersky, Wang, and Zhang}{Zamani
  et~al\mbox{.}}{2017}]%
        {Zamani:2017}
\bibfield{author}{\bibinfo{person}{Hamed Zamani}, \bibinfo{person}{Michael
  Bendersky}, \bibinfo{person}{Xuanhui Wang}, {and} \bibinfo{person}{Mingyang
  Zhang}.} \bibinfo{year}{2017}\natexlab{}.
\newblock \showarticletitle{Situational Context for Ranking in Personal
  Search}. In \bibinfo{booktitle}{{\em WWW '17}}. \bibinfo{pages}{1531--1540}.
\newblock


\bibitem[\protect\citeauthoryear{Zamani and Croft}{Zamani and Croft}{2016a}]%
        {Zamani:2016:ICTIR:emb}
\bibfield{author}{\bibinfo{person}{Hamed Zamani} {and}
  \bibinfo{person}{W.~Bruce Croft}.} \bibinfo{year}{2016}\natexlab{a}.
\newblock \showarticletitle{{Embedding-based Query Language Models}}. In
  \bibinfo{booktitle}{{\em ICTIR '16}}. \bibinfo{pages}{147--156}.
\newblock


\bibitem[\protect\citeauthoryear{Zamani and Croft}{Zamani and Croft}{2016b}]%
        {Zamani:2016:ICTIR:pqv}
\bibfield{author}{\bibinfo{person}{Hamed Zamani} {and}
  \bibinfo{person}{W.~Bruce Croft}.} \bibinfo{year}{2016}\natexlab{b}.
\newblock \showarticletitle{{Estimating Embedding Vectors for Queries}}. In
  \bibinfo{booktitle}{{\em ICTIR '16}}. \bibinfo{pages}{123--132}.
\newblock


\bibitem[\protect\citeauthoryear{Zamani, Dadashkarimi, Shakery, and
  Croft}{Zamani et~al\mbox{.}}{2016}]%
        {Zamani:2016:CIKM}
\bibfield{author}{\bibinfo{person}{Hamed Zamani}, \bibinfo{person}{Javid
  Dadashkarimi}, \bibinfo{person}{Azadeh Shakery}, {and}
  \bibinfo{person}{W.~Bruce Croft}.} \bibinfo{year}{2016}\natexlab{}.
\newblock \showarticletitle{{Pseudo-Relevance Feedback Based on Matrix
  Factorization}}. In \bibinfo{booktitle}{{\em CIKM '16}}.
  \bibinfo{pages}{1483--1492}.
\newblock


\bibitem[\protect\citeauthoryear{Zhai, Cohen, and Lafferty}{Zhai
  et~al\mbox{.}}{2003}]%
        {Zhai:2003}
\bibfield{author}{\bibinfo{person}{ChengXiang Zhai},
  \bibinfo{person}{William~W. Cohen}, {and} \bibinfo{person}{John Lafferty}.}
  \bibinfo{year}{2003}\natexlab{}.
\newblock \showarticletitle{Beyond Independent Relevance: Methods and
  Evaluation Metrics for Subtopic Retrieval}. In \bibinfo{booktitle}{{\em SIGIR
  '03}}. \bibinfo{pages}{10--17}.
\newblock


\bibitem[\protect\citeauthoryear{Zhai and Lafferty}{Zhai and Lafferty}{2001}]%
        {Zhai:2001}
\bibfield{author}{\bibinfo{person}{Chengxiang Zhai} {and} \bibinfo{person}{John
  Lafferty}.} \bibinfo{year}{2001}\natexlab{}.
\newblock \showarticletitle{{Model-based Feedback in the Language Modeling
  Approach to Information Retrieval}}. In \bibinfo{booktitle}{{\em CIKM '01}}.
  \bibinfo{pages}{403--410}.
\newblock


\bibitem[\protect\citeauthoryear{Zhai and Lafferty}{Zhai and Lafferty}{2004}]%
        {Zhai:2004}
\bibfield{author}{\bibinfo{person}{Chengxiang Zhai} {and} \bibinfo{person}{John
  Lafferty}.} \bibinfo{year}{2004}\natexlab{}.
\newblock \showarticletitle{{A Study of Smoothing Methods for Language Models
  Applied to Information Retrieval}}.
\newblock \bibinfo{journal}{{\em ACM Trans. Inf. Syst.\/}}
  \bibinfo{volume}{22}, \bibinfo{number}{2} (\bibinfo{year}{2004}),
  \bibinfo{pages}{179--214}.
\newblock


\bibitem[\protect\citeauthoryear{Zheng and Callan}{Zheng and Callan}{2015}]%
        {Zheng:2015}
\bibfield{author}{\bibinfo{person}{Guoqing Zheng} {and} \bibinfo{person}{Jamie
  Callan}.} \bibinfo{year}{2015}\natexlab{}.
\newblock \showarticletitle{{Learning to Reweight Terms with Distributed
  Representations}}. In \bibinfo{booktitle}{{\em SIGIR '15}}.
  \bibinfo{pages}{575--584}.
\newblock


\bibitem[\protect\citeauthoryear{Zhou, He, Zhao, and Hu}{Zhou
  et~al\mbox{.}}{2015}]%
        {Zhou:2015}
\bibfield{author}{\bibinfo{person}{Guangyou Zhou}, \bibinfo{person}{Tingting
  He}, \bibinfo{person}{Jun Zhao}, {and} \bibinfo{person}{Po Hu}.}
  \bibinfo{year}{2015}\natexlab{}.
\newblock \showarticletitle{{Learning Continuous Word Embedding with Metadata
  for Question Retrieval in Community Question Answering}}. In
  \bibinfo{booktitle}{{\em ACL '15}}. \bibinfo{pages}{250--259}.
\newblock


\end{thebibliography}



\end{document}